\pdfoutput=1
 
\documentclass[cernpreprint, atlasdraft=false, texlive=2023, UKenglish, texmf, orcidlogo]{atlasdoc}

\DeclareBoolOption[true]{LANGEDIT}

\usepackage{atlaspackage}

\makeatletter
\DeclareOldFontCommand{\rm}{\normalfont\rmfamily}{\mathrm}
\DeclareOldFontCommand{\sf}{\normalfont\sffamily}{\mathsf}
\DeclareOldFontCommand{\tt}{\normalfont\ttfamily}{\mathtt}
\DeclareOldFontCommand{\bf}{\normalfont\bfseries}{\mathbf}
\DeclareOldFontCommand{\it}{\normalfont\itshape}{\mathit}
\DeclareOldFontCommand{\sl}{\normalfont\slshape}{\@nomath\sl}
\DeclareOldFontCommand{\sc}{\normalfont\scshape}{\@nomath\sc}
\makeatother
 
\usepackage{atlasbiblatex}
 
\usepackage{atlasphysics}

\graphicspath{{logos/}{figures/}}
 
\usepackage{ANA-EXOT-2020-15-PAPER-defs}

\usepackage{graphicx}
\AtlasTitle{
Search for new phenomena in multi-body invariant masses in events with at least one isolated
lepton and two jets using $\sqrt{s} = 13$~TeV proton--proton collision  data collected by the ATLAS detector
}
 
\PreprintIdNumber{CERN-EP-2022-179}
\AtlasJournalRef{JHEP 07 (2023) 202}
\AtlasDOI{DOI:10.1007/JHEP07(2023)202}
 
\AtlasAbstract{
A search for resonances in events with at least one isolated
lepton ($e$ or $\mu$)  and two jets is performed using $139~{\text{fb}}^{-1}$ of $\sqrt{s}=13$~TeV
proton--proton collision data recorded by the ATLAS detector at the LHC.
Deviations from a smoothly falling background hypothesis are tested in three- and four-body invariant mass distributions constructed from leptons and jets, including
jets identified as originating from bottom quarks.
Model-independent limits on generic resonances characterised by cascade
decays of particles leading to multiple jets and leptons in the final state are presented.
The limits are calculated using Gaussian shapes with different widths for the invariant masses.
The multi-body invariant masses are also used to set 95\% confidence level upper limits on
the cross-section times branching ratios for the production and subsequent decay of
resonances predicted by several new physics scenarios.
}
 
\author{The ATLAS Collaboration}
 
\AtlasRefCode{EXOT-2020-15}

\AtlasJournal{JHEP}

\hypersetup{pdftitle={ATLAS document},pdfauthor={The ATLAS Collaboration}}

\begin{document}
 
\maketitle

 
\section{Introduction}
 
Searches for resonances in dijet invariant mass distributions provide a means to investigate a wide range of theories beyond the Standard Model (BSM theories).
Such searches are sensitive to heavy particles that decay into two partons (quarks or gluons) which, following hadronisation, form jets (i.e.\ dijets).
Previous searches for heavy resonances in dijet mass distributions performed by ATLAS~\cite{2020dijets,Aaboud:2019zxd,Aaboud:2018fzt} and
CMS~\cite{CMS:2019gwf,CMS:2022zoc,CMS:2019mcu} have set limits on the production of heavy resonances over a range of masses for a variety of BSM models.
 
Searches in dijet mass distributions were also performed in events triggered by leptons \cite{Aad:2020kep}. In contrast to the searches in inclusive dijet events, an exclusive selection requiring events with a light lepton (electron $e$ or muon $\mu$, referred to simply as a `lepton' herein) in addition to dijets in the final state increases the sensitivity to various exotic models. Using lepton triggers to select these events
provides a means to study low-mass regions that are difficult to access in inclusive dijet events, where the large jet production rate imposes
high threshold requirements on jet triggers.
In addition, requiring an additional lepton reduces the QCD multi-jet background rate and thus can be favourable for BSM model searches. A recent ATLAS search \cite{ATLAS:2022nmi} similarly looked for resonances in complementary final states with $Z \to ll + X$ events.
 
The searches \cite{Aad:2020kep} for a heavy dijet resonance in two-jet events triggered by leptons can be extended into a search for a resonance in a final state which includes two jets ($jj$) and a lepton ($\ell$). Searches for narrow peaks in three-body decays (such as invariant masses of two jets and a lepton, $\mjjl$) are sensitive to a new class of models with cascade decays of heavy particles which have not been studied extensively at the LHC.
Such studies in multi-body decays can be performed in exactly the same way as searches for resonances in dijet mass ($\mjj$) distributions using a data-driven technique for background determination \cite{Chekanov:2021huv}. This technique mitigates problems arising from insufficient theoretical knowledge of the Standard Model  (SM) background in Monte Carlo (MC) simulations.
 
There are a number of theoretical models~\cite{Abercrombie:2015wmb,AltarelliMele,Roitgrund_2021, PhysRevD.98.035010,Agashe:2017wss,Agashe:2016kfr} that, for a certain choices of parameters,
can exhibit resonance-like structures in multi-body invariant masses.
For example \cite{Chekanov:2021huv}, the invariant mass $\mjjl$ for decays  such as $X\to Y\> W\to jj \ell\nu$ (with a $W$ boson decaying to $\ell\nu$) can be used in searches for a heavy unknown particle $X$ in scenarios where the mass-splitting between $X$ and a second unknown particle $Y$ is not large (a few hundred \GeV). This leads to a relatively small transverse momentum for the neutrino. Therefore, the inclusion of missing transverse momentum may not improve the reconstruction of $X$ masses.
 
More generally, a  generic $s$-channel process, $X\rightarrow Y\> C$, can be considered, where $X$ is an unknown heavy  particle. The particles $Y$ and $C$ are SM or BSM particles that produce
quarks, gluons, leptons or other particles via two-body decay kinematics.
As a result, an enhancement in multi-body invariant masses is expected near the mass of the particle $X$, which can be used in searches.
However, two-body invariant masses may not be the best tool for identifying such events if $Y$ and $C$ are broad resonances.
The experimental limitations due to large widths, $\Gamma/m>0.15$ (where $\Gamma$ is the width and $m$ is the mass of a resonance), of dijet resonances are discussed in the previous dijet searches~\cite{2020dijets,Aaboud:2019zxd,Aaboud:2018fzt,Aad:2020kep}.
 
This study reports model-independent searches for new phenomena
in three- and four-body invariant mass distributions in events  with at least one isolated lepton. Such searches can lead to the discovery of heavy resonances with cascade decays in situations where two-body decays are broad, or affected by a large SM background. Three- and four-body masses
have not been explored extensively at the LHC using data-driven techniques for background determination.
These invariant masses are also used here to study  a number of specific BSM models,
such as dark-matter models with an axial-vector mediator \cite{Abercrombie:2015wmb},  the Sequential Standard Model with $W'$ and $Z'$ bosons \cite{AltarelliMele},
composite resonances breaking lepton-flavour universality \cite{PhysRevD.98.035010},
and radion models \cite{Agashe:2017wss,Agashe:2016kfr}.
The search uses an integrated luminosity of $\lumi~\text{fb}^{-1}$ of $\sqrt{s}=13$~\TeV\
proton--proton collision data recorded by the ATLAS detector.


\section{ATLAS detector}
The ATLAS detector~\cite{PERF-2007-01,ATLAS-TDR-2010-19,PIX-2018-001} consists of an inner tracking detector
(ID), surrounded by
a superconducting solenoid which provides a 2\,T magnetic field, electromagnetic and hadronic
calorimeters, and a muon spectrometer (MS).
The ID provides tracking in the
pseudorapidity\footnote{ATLAS uses a right-handed coordinate system with its origin at the nominal
interaction point (IP) in the centre of the detector and the $z$-axis along the beam line.
The
$x$-axis points from the IP to the centre of the LHC ring, and the $y$-axis points upwards.
Cylindrical coordinates $(r,\phi)$ are used in the transverse plane, $\phi$ being the azimuthal
angle around the beam line.
Transverse momentum and energy are defined as $p_\text{T}=p\sin
\theta$ and $E_\text{T}=E \sin \theta$, respectively. The pseudorapidity is defined in terms of
the polar angle $\theta$ as $\eta = -\ln \tan(\theta/2)$. The angular separation between
two objects in $\eta$--$\phi$ space is defined as $\Delta R = \sqrt{(\Delta \eta)^2 + (\Delta\phi)^2}$.}
region $|\eta| \le 2.5$ and consists of
silicon pixel and microstrip detectors surrounded by a transition radiation tracker, which also provides
information for electron identification.
Each tracking detector consists of a central barrel and two endcap sections.
 
The electromagnetic calorimeter is a sampling device made of lead absorbers with liquid argon
(LAr) as active medium.
It comprises a barrel ($|\eta| \le 1.475$) and two endcaps ($1.375 \le |\eta| \le 3.2$).
To facilitate corrections for energy losses upstream of the
calorimeter, the cryostat is equipped with a presampler layer in the region $|\eta| \le 1.8$.
Hadronic sampling calorimetry is provided by a steel and scintillator-tile calorimeter in the
region $|\eta| \le 1.7$, complemented by a copper/LAr system in the region $1.5 \le|\eta|\le3.2$.
The forward region ($3.1 \le |\eta| \le 4.9$) is equipped with both electromagnetic and hadronic
calorimeters composed of copper/LAr and tungsten/LAr, respectively.
 
The muon spectrometer is the outermost ATLAS subsystem.
It detects muons in the pseudorapidity region up to $|\eta| = 2.7$, with triggering capability up to $|\eta| = 2.4$.
The spectrometer consists of a barrel ($|\eta| \le 1.05$) and two endcap
sections ($1.05 \le |\eta| \le 2.7$). A system of three large superconducting air-core toroidal magnets,
each with eight coils, provides a magnetic field with a bending integral of about 2.5~Tm in
the barrel and up to 6~Tm in the endcaps.
 
The trigger system~\cite{Aaboud:2016leb} consists of a first-level trigger implemented in hardware
using a subset of the detector information to
accept events from the 40~MHz bunch crossings at a rate of 100~kHz,
followed by a
software-based trigger implemented in a large computer farm,
which reduces the acceptance rate so that events are recorded at about 1~kHz.
An extensive software suite~\cite{ATL-SOFT-PUB-2021-001} is used in data simulation, in the reconstruction and
analysis of real and simulated data, in detector operations, and in the trigger and data acquisition systems of the experiment.


 
\section{Monte Carlo simulations}
\label{sec:mc}
 
Monte Carlo (MC) simulations of various SM background processes are used to estimate their
contributions to the multi-body mass distributions and
to validate some aspects of the analysis procedure. The considered sources of SM background are the QCD multi-jet, $t\bar{t}$, single-top and $W$+jets processes.
The QCD multi-jet event sample was created at  leading order in QCD
with parton showering using the \Pythia[8]~\cite{Sjostrand:2007gs} generator. It uses  the \NNPDF[2.3]~\cite{Ball:2012cx} set of parton distribution functions (PDF) and a set of tuned
parameters called the A14 tune~\cite{ATL-PHYS-PUB-2014-021}.
The $t\bar{t}$, single-top ($t$- and $Wt$-channels) and $W$+jets events were produced at next-to-leading order in QCD using the \POWHEGBOX[v2]~\cite{Nason:2004rx,Frixione:2007vw,Alioli:2010xd,Frixione:2007nw} generator interfaced with \Pythia[8].
This simulation uses the \CT[10nlo] PDF set~\cite{Lai:2010vv} and the \AZNLO tune~\cite{Aad:2014xaa}.
 
\begin{figure}
\begin{center}
 
\subfloat[]{
\includegraphics[width=0.45\textwidth]{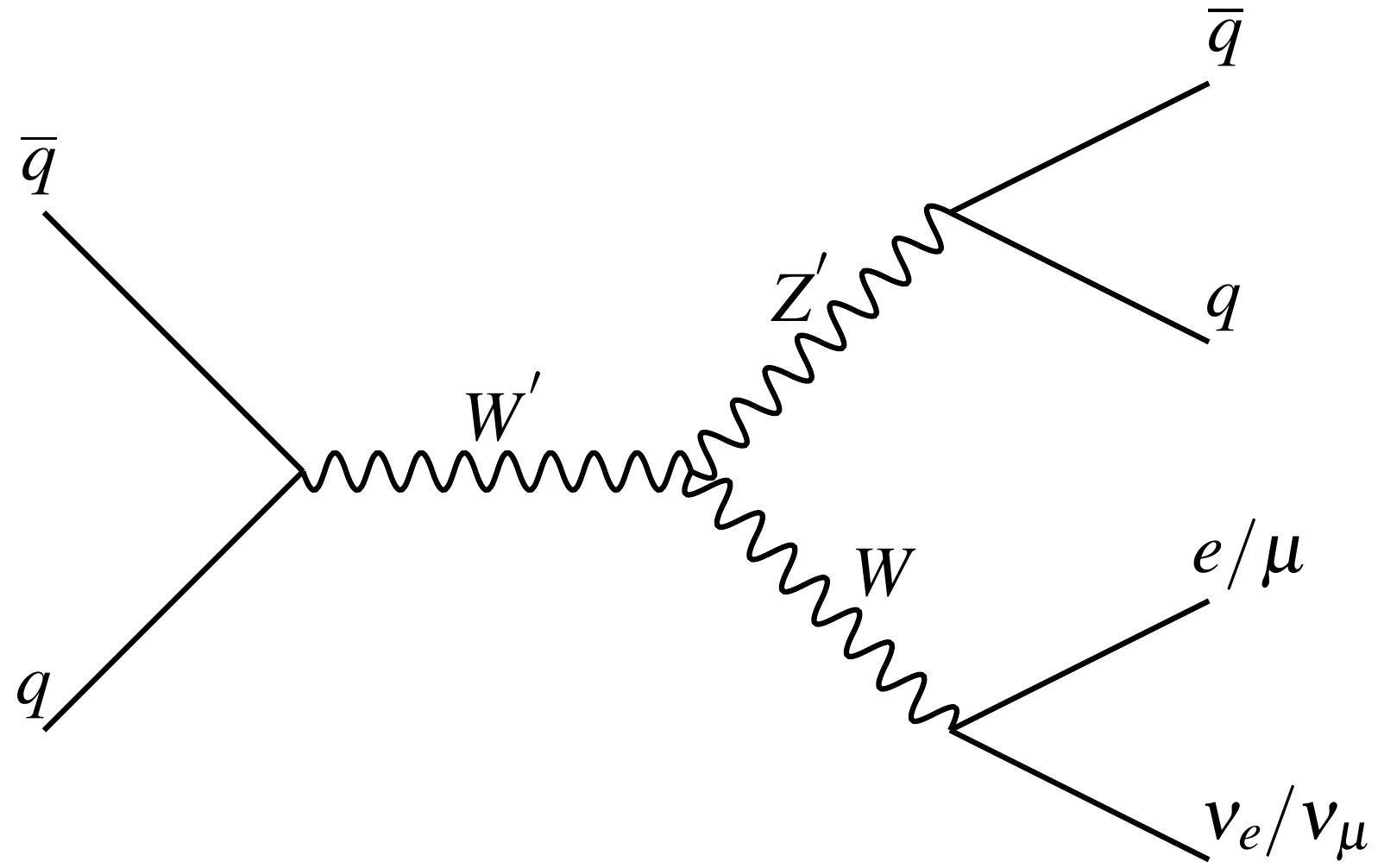}\hfill
}
\subfloat[]{
\includegraphics[width=0.45\textwidth]{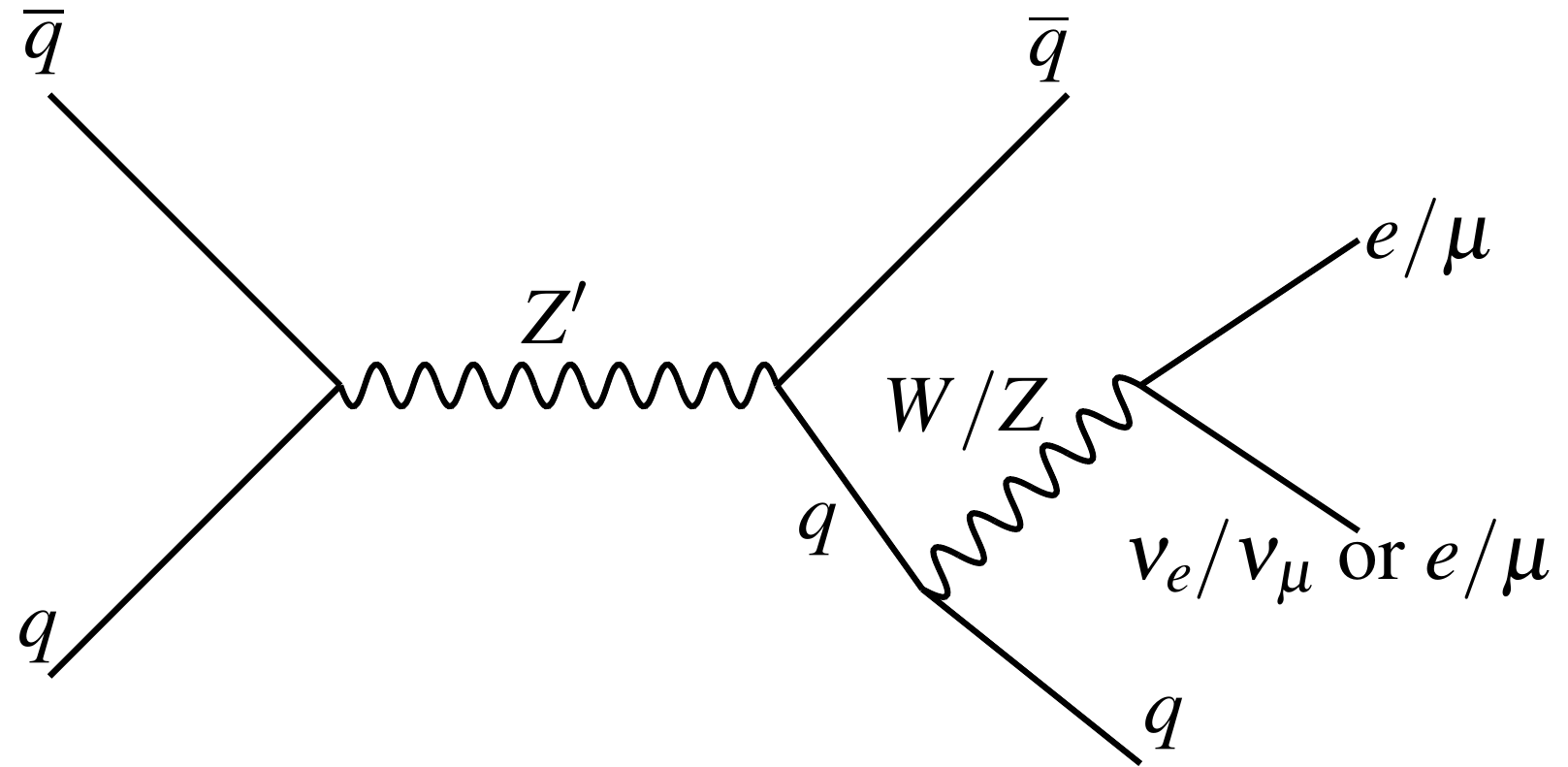}\hfill
}
 
\subfloat[]{
\includegraphics[width=0.45\textwidth]{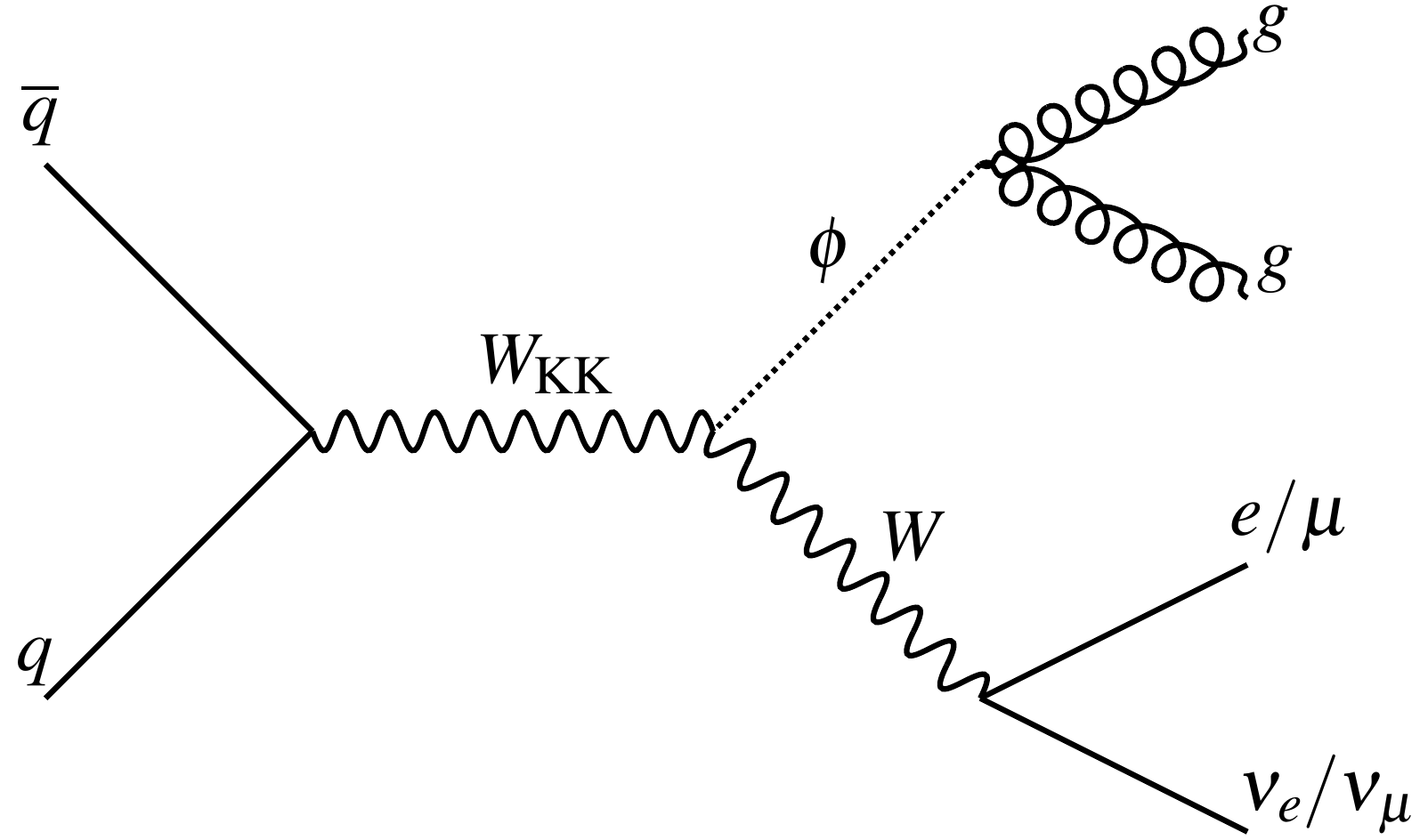}\hfill
}
\subfloat[]{
\includegraphics[width=0.45\textwidth]{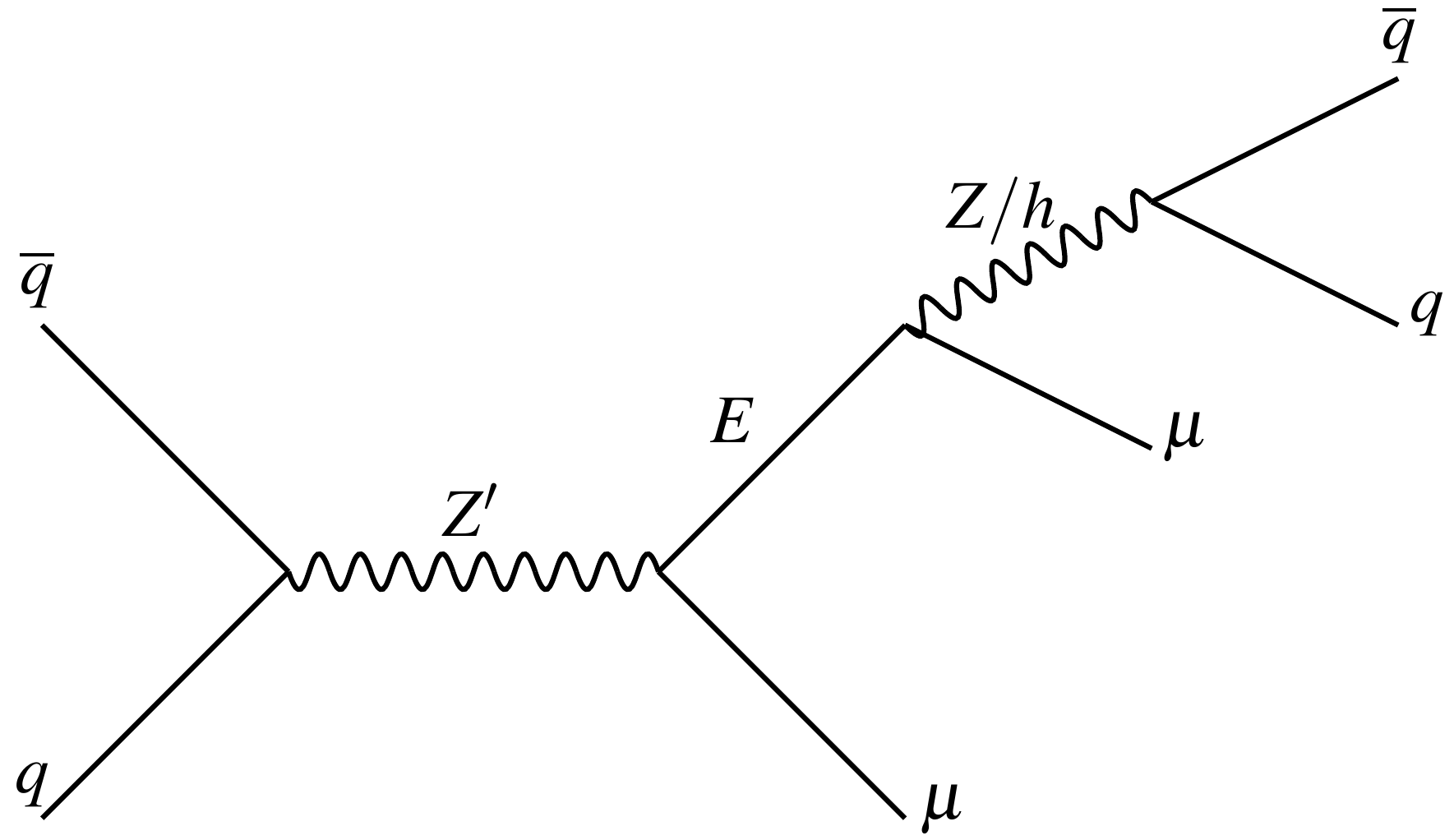}\hfill
}

\end{center}
\caption{
Representative Feynman diagrams for the models considered in this analysis: (a) $W' \to Z' \>  W$ production
in the SSM, (b) the simplified DM model,  (c) the radion model, (d) the composite-lepton model.  The text provides more detailed descriptions.
}
\label{fig:feynman}
\end{figure}

For the model-dependent searches, MC simulations are also used to predict the expected signal shapes for the BSM models mentioned earlier:
(1) $W'\to W Z'$ in the Sequential Standard Model (SSM),
(2) a simplified dark-matter (DM) model with an axial-vector mediator $Z'$,
(3) a Kaluza--Klein (KK) gauge boson model with a SM $W$ boson and a radion that decays into two gluons,
(4) a model with composite SU(2)$_\text{L}$ fermion doublets that breaks lepton-flavour universality (`composite lepton').
Representative Feynman diagrams for these models are shown in Figure~\ref{fig:feynman}.
 
Simulations for the SSM \cite{AltarelliMele} were performed at leading order in QCD using \Pythia[8] with
the \CT[10nlo] PDF set. Event-level simulations for all other BSM models were performed with \MGNLO \cite{Alwall:2014hca} at leading order in QCD. The model assumes the heavy $Z'$ boson decays into $q$ and $\bar{q}$, which give rise to two high-energy jets, while the $W$ boson decays leptonically. 
The $W'\to W Z'$ branching fraction was chosen to be 0.5 and the mass difference between the $W'$ and $Z'$ bosons was set to 250~\GeV.
The latter setting yields the largest predicted cross-section for the desired final state.
This model is also used to estimate systematic uncertainties for the limits when using generic signals approximated by Gaussian functions.
 
A benchmark simplified DM model \cite{Abercrombie:2015wmb} includes an axial-vector mediator $Z'$ which decays as $q\bar{q}\rightarrow Z' \rightarrow q\bar{q}$, and one of the quarks radiates a $W$ or $Z$ boson which then decays via $W/Z \rightarrow \ell \nu / \ell \ell$.
This model assumes the leptophobic couplings $g_q = 0.25$, $g_\ell = 0$, and $g_\text{DM} = 1$, following the recommendations
of the LHC Dark Matter Working Group~\cite{Abercrombie:2015wmb}.
Here $g_q$, $g_\ell$ and $g_\text{DM}$ are the couplings of the mediator to quarks, leptons and the DM particle, respectively.
 
A warped-extra-dimensions model assumes KK excitations of gauge bosons and their decay into a radion and a SM gauge boson, e.g.\ $q\bar{q} \rightarrow W_\text{KK} \rightarrow \phi W$, followed by $W\rightarrow \ell \nu$ and the radion $\phi$ decaying into a pair of gluons \cite{Agashe:2017wss,Agashe:2016kfr}.
A mass difference between KK bosons and radions of 250~\GeV\ was chosen for all samples in order to minimise signal peak widths. A larger mass difference leads to the production of higher-\pT neutrinos that need to be taken into account in the $W_\text{KK}$ mass reconstruction.
 
The composite-lepton model \cite{PhysRevD.98.035010} predicts a $Z'$-like vector boson that decays into a new vector-like composite lepton $E$ along with a SM lepton, $Z' \rightarrow E \ell$. The composite lepton $E$ can then decay as $E \rightarrow Z/h \> \ell$, with the $Z/h$ boson eventually decaying into two jets. The two leptons produced in this process have sufficient transverse momenta to give higher  acceptances in the $\mjjll$ channel.

The parton showering and hadronisation processes for all the used BSM models were performed using \Pythia[8] with the A14 tune. All SM background MC processes were passed through the full ATLAS detector simulation~\cite{Aad:2010ah} based on \GEANT~\cite{Agostinelli:2002hh};
samples for the BSM signal processes were simulated using the ATLAS fast simulation framework,
ATLFAST-II, which uses parameterisations of electromagnetic and hadronic
showers in the calorimeters. Additional simulated $pp$ collisions generated using \Pythia[8],
with the A3 set of tuned parameters~\cite{ATL-PHYS-PUB-2016-017} and the \NNPDF[2.3] PDF set,
were overlaid to simulate the effects of pile-up in a manner that matches the multiplicity distribution of additional collisions in the data. Simulated MC events are reconstructed and analysed with the same algorithms as used for data.
All simulated events are corrected so that the object identification, reconstruction and trigger efficiencies, energy scales and energy resolution match those determined from data control samples.
After these corrections, additional systematic uncertainties are applied to cover differences between the data and MC simulations as described in Section~\ref{sec:sys}.
 
\section{Object definitions and event selection}
 
The analysis presented in this paper is based on data collected with the ATLAS detector during the
2015--2018 data-taking period, referred to as Run~2.
The data sample corresponds to an integrated luminosity of $\lumi$~fb$^{-1}$
with an uncertainty of 1.7\%.
The uncertainty was derived from a calibration of the luminosity scale
using $x$--$y$ beam-separation scans, following a methodology similar to that detailed in Ref.~\cite{ATLAS-CONF-2019-021} and using data from the LUCID-2 detector~\cite{LLUCID2} for the baseline measurement.
 
Candidate events were accepted by either single-muon or single-electron triggers~\cite{Aaboud:2016leb,ATLAS:2019dpa,ATLAS:2020gty}
with various thresholds for the transverse momentum $p_\text{T}$ of muons or transverse energy $E_\text{T}$ of electrons,
as well as data quality and lepton isolation requirements.
The lowest $p_\text{T}$ ($E_\text{T}$)
threshold without trigger prescaling is 24 (26)~\GeV\ and includes a lepton isolation
requirement that is not applied for triggers with higher thresholds.
A trigger matching requirement \cite{Aaboud:2016leb} is applied to ensure that
the reconstructed lepton lies in the vicinity of the corresponding trigger-level object.
 
Muons  reconstructed from a combination of hits in the ID and MS
are required to have $p_\text{T}\ge 7$~\GeV\ and $|\eta|\le 2.5$. They must satisfy `medium' quality criteria~\cite{Aad:2016jkr}.
Identification requirements based on the number of hits
in the ID and MS subsystems, as well as the significance of the difference
$|q/p_\text{MS} - q/p_\text{ID}|$, where $q$ is the charge and $p_\text{MS}$ ($p_\text{ID}$)
is the momentum as measured in the MS (ID), are applied to the combined tracks.
Muon tracks are required to satisfy $|d_0|/\sigma(d_{0}) \le 3$ and $|\zzsth| \le 0.5$~mm,
where $d_0$ is the transverse impact parameter relative to the beam line, $\sigma(d_0)$
is its uncertainty, $z_0$ is the distance along the beam line to the primary vertex
from the point where $d_0$ is measured, and $\theta$ is the polar angle of the track.
The primary vertex is chosen as the vertex with the highest $\sum{\pt^2}$, where
the sum is over tracks associated with that vertex and having $\pt>500$~\MeV; at least two such tracks are required.
 
Electrons are identified as energy clusters in the electromagnetic calorimeter~\cite{Aad:2019tso} 
matched to tracks in the ID, with requirements of $E_\text{T} > 7$~\GeV\ and $|\eta|\le 2.47$.
Candidate electrons must have a track satisfying $|d_0|/\sigma(d_{0}) \le 5$ and
$|\zzsth| \le 0.5$~mm and must meet `tight' quality criteria~\cite{Aad:2019tso}.  
Electrons within the barrel--endcap transition region of the electromagnetic calorimeter, $1.37 \le |\eta| \le 1.52$,
or which share a track with an identified muon, are discarded.
 
Leptons are also required to be isolated from other objects in the event, using
$\pt$-dependent criteria based on calorimeter and tracking information.
The isolation parameters of the FCTight isolation working point
\cite{ATLAS:2019dpa} were  tuned to provide a constant efficiency as a function of transverse momentum,
and the highest background rejection below 60~\GeV. 
The lepton isolation and $\pt$ requirements allow a consistent definition of lepton candidates when considering
data-taking periods in which different trigger configurations were used.
Lepton misidentification rates for the chosen isolation requirements are discussed in Refs.~\cite{Aad:2016jkr,Aad:2019tso}.
The trigger efficiencies for electrons and muons are above 90\% \cite{Hrynova:2668252,ATLAS:2020auj}.

Jets are reconstructed using the anti-$k_t$ algorithm~\cite{Cacciari:2008gp} with a radius parameter of
$R=0.4$, as implemented in the FastJet package~\cite{fastjet}, using particle-flow objects \cite{ATLAS:2017ghe}
as inputs.
These jets are corrected for contributions arising from additional collisions in the same or neighbouring bunch
crossings (pile-up)~\cite{ATLAS-CONF-2014-018}, and calibrated to the particle energy scale (i.e.\ before interaction with the detector)~\cite{ATLAS:2020cli}.
Jet candidates must have $p_\text{T}\ge 30$~\GeV\ and be within $|\eta|\le 2.4$.
To suppress jets arising from pile-up, a jet-vertex-tagging technique~\cite{Aad:2015ina,ATL-PHYS-PUB-2019-026} is
applied, using the "Tight" working point for central jets with $|\eta| < 2.5$. Jets originating from bottom quarks (`$b$-jets') are identified  using the DL1r algorithm \cite{ATLAS:2019bwq,ATL-PHYS-PUB-2020-009} with an offline working point providing 77\% $b$-jet tagging efficiency \cite{DL1R}.
 
An overlap-removal procedure is applied to remove ambiguities which arise when the same object is reconstructed by different algorithms.  Candidate jets with fewer than three
associated tracks are discarded if they lie within a cone of $\Delta R=0.2$ around a muon candidate. Jets with any number of associated tracks
are discarded if they are within $\Delta R = 0.2$ of an electron.
Electron and muon candidates are discarded if they are within a cone of $\Delta R=0.4$ around the axis of any remaining jets.
 
Following the lepton and jet selection,
a signal region is defined by requiring at least one isolated lepton ($e$ or $\mu$)
with $p_\text{T}^\ell \ge 60$~\GeV\ and at least two jets.
The invariant masses are constructed by combining the two jets having the highest $p_\text{T}$ with one ($\mjjl$) or two ($\mjjll$) leptons with the highest $p_\text{T}^\ell$. No requirements  on the charges of leptons for $\mjjll$ are applied, as this keeps
the selection as model-independent as possible.
Invariant masses with at least one $b$-jet ($\mjbl$) or two $b$-jets ($\mbbl$) are also
reconstructed. These invariant masses are sensitive to various BSM scenarios including a few supersymmetric models with R-parity-violating interactions \cite{Das:2017kfo,Earl:2018snx}.
 
For all distributions,
only events with invariant masses above $0.4$~\TeV\ are considered;
this minimum value is chosen in order to avoid
the low-mass region where the event rate does not
decrease monotonically with increasing mass, due to a kinematic bias in jet $\pt$ from the minimum-$p_\text{T}^\ell$ requirement imposed on leptons.
 
\section{Analysis procedure}
\label{sec:ana}
 
A new resonant state $X$  with a mass $\mX$ can decay
into partons and leptons, either directly
or via complex cascade decays. Such decays could lead to an observable excess of events in multi-body invariant masses,
such as $\mjjl$ and $\mjjll$, near $\mX$ on an otherwise smooth and monotonically decreasing background distribution.
This analysis presents a search for such an excess in the mass range above $0.4$~\TeV\ for $\mjjl$, $\mjjll$, $\mjbl$ and $\mbbl$. These signal regions are not orthogonal, with $\mjjl$ being the most inclusive region, and all other regions being subsets.
 
The bin widths of the multi-body distributions are chosen to be larger than the resolution for the four-body mass $\mjjll$, which is fully dominated by the two-jet mass resolution at a given mass.
The bin widths are increased from 19~\GeV\ to 148~\GeV\ over the range of 0.4--8~\TeV\
to reflect the changes in the mass resolution.
The following fit function is used to model the shape of the estimated background:
\begin{equation}
f(x) = p_1 (1 - x)^{p_2} x^{p_3 + p_4\ln x + p_5 \ln^2 x},
\label{eq:function}
\end{equation}
where $x \equiv m /\sqrt{s}$ and the $p_i$\ ($i=1,\cdots 5$) are five free parameters. In the following, this function is referred to as a $5p$ function.The 4-parameter fit function with $p_5=0$ describes dijet mass spectra of Monte Carlo simulations ~\cite{ATLAS:2010ivc}, while the 5-parameter fit function is a natural extension used in recent publications~\cite{Aad:2014aqa,Aaboud:2018fzt,Aaboud:2019zxd,Aad:2020kep}.
 
According to MC simulations, about 90\% of events contributing to the $\mjjl$ invariant mass spectrum originate from the multi-jet background where a jet is misidentified as an electron. The study of such events is limited by the available MC statistics.
To investigate the ability of Eq.~(\ref{eq:function}) to accurately describe the
background in the signal region, a likelihood fit is performed on a background-enhanced sample obtained from a `loose electron' control region (LE-CR) defined in data. This control region is dominated by multi-jet events where jets are misidentified as electrons.
The LE-CR is populated by selecting events that contain at least two jets and one electron, where the electron must satisfy only loose identification criteria, while it must fail the more stringent identification criteria. This ensures the orthogonality of this control region with the signal region that requires the more stringent criteria for electron selection.

The main purpose of the LE-CR is to investigate possible effects related to jet reconstruction, or to leptons misidentified as jets, which may lead to structures in the mass distributions that  could be misinterpreted as a potential signal~\cite{Aad:2020kep}. The LE-CR does not accurately represent the signal region since it misses the important contributions from W/t production. The LE-CR distributions are created for each signal region. The number of LE-CR events is a factor of 10 (40) smaller than the number of events in the ($\mjjl$) ($\mjjll$) signal regions. The $5p$ fit describes the LE-CR well, without structures in the pulls of the fit that could be interpreted as signals.
 
\begin{figure}
\begin{center}
 
\subfloat[$\mjjl$]{
\includegraphics[width=0.43\textwidth]{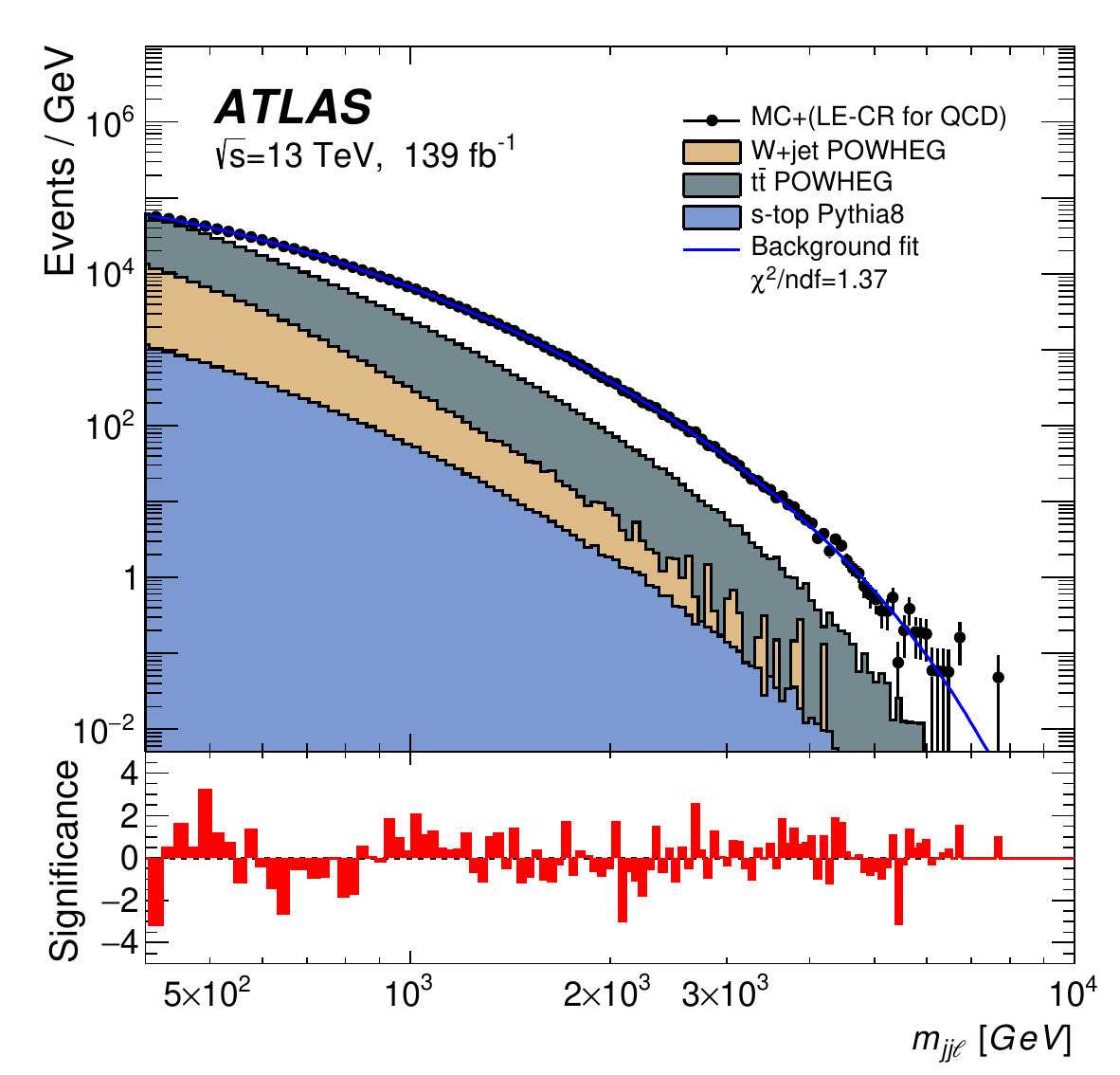}\hfill
}
\subfloat[$\mjjll$]{
\includegraphics[width=0.43\textwidth]{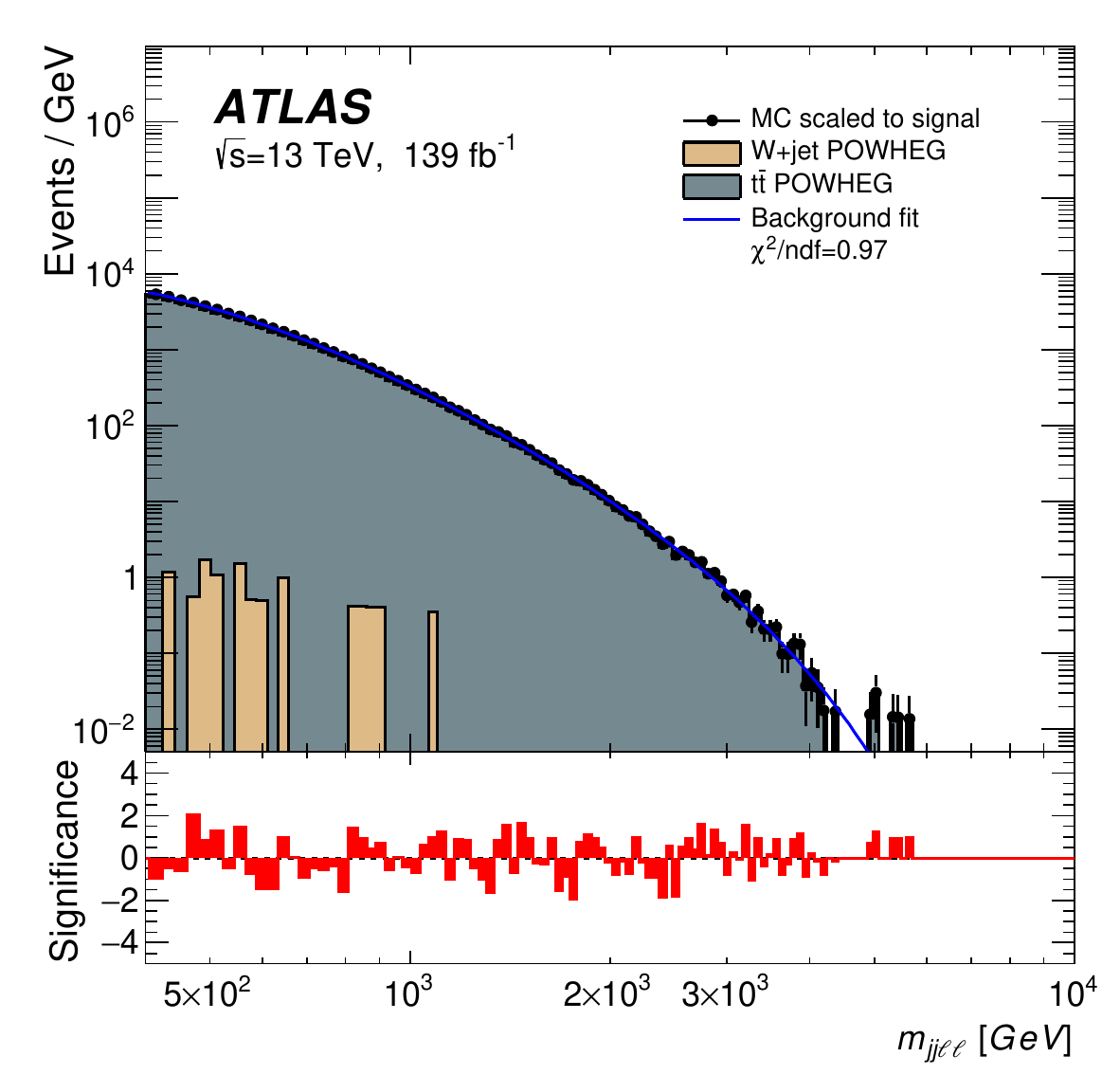}\hfill
}
 
\subfloat[$\mjbl$]{
\includegraphics[width=0.43\textwidth]{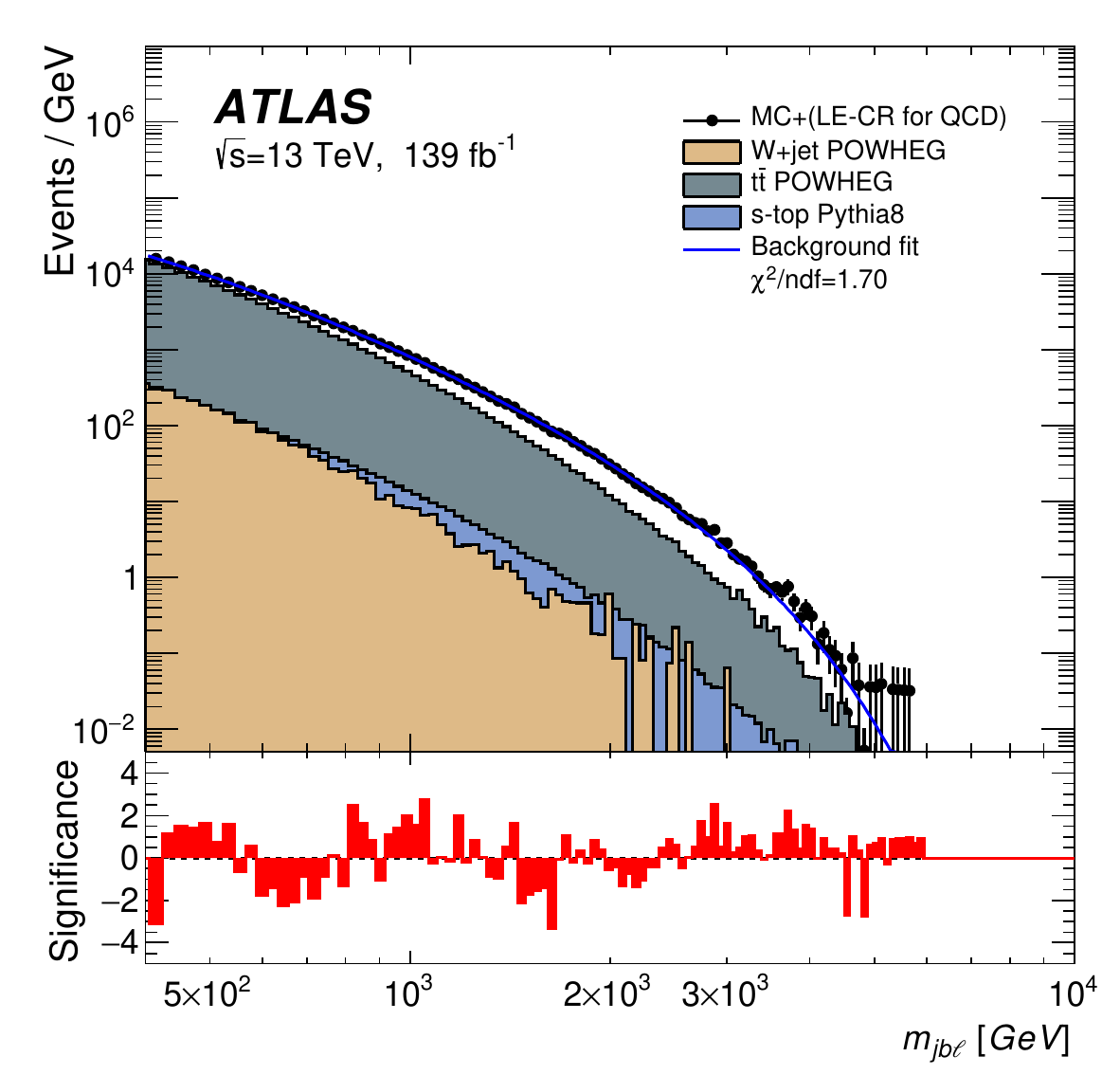}\hfill
}
\subfloat[$\mbbl$]{
\includegraphics[width=0.43\textwidth]{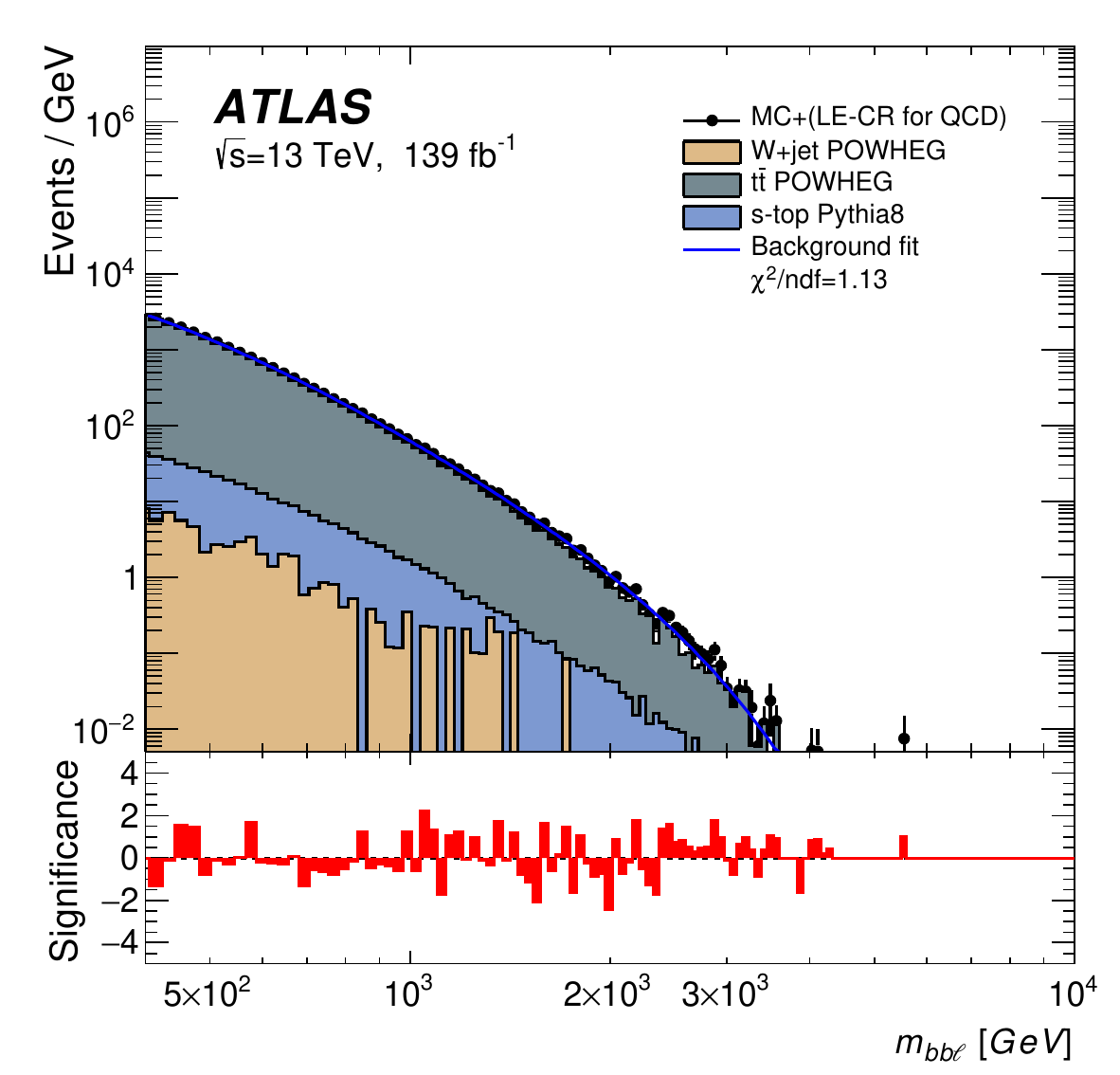}\hfill
}
 
\end{center}
\caption{Invariant mass distributions (a) $\mjjl$, (b) $\mjjll$, (c) $\mjbl$ and (d) $\mbbl$ with the $5p$ background fits in the MC+LE-CR control region. The lower panels show the bin-by-bin significances of deviations from the background hypothesis.  The solid symbols show the sum of Monte Carlo contributions and loose electron control region. Only the three main MC contributions are shown (overlaid): (1) W boson plus jet production; (2) $t\bar{t}$ production; and (3) single-$t$ production.
}
\label{fig:mjjl:signal:mc_with_fit}
\end{figure}

Due to the over-representation of QCD multi-jet events in the LE-CR, $W+$jets, $t\bar{t}$ and single-top MC events were added to the LE-CR using the predicted cross-sections. The normalisations of the simulated distributions were set to the expected rate in each signal region to create a hybrid control region (MC+LE-CR).  The combined contribution from $W$ boson and top quark processes in the MC simulations
varies from 10\% to 95\% as a function of invariant mass and the $b$-jet selection.
The fit hypothesis of Eq.~(\ref{eq:function}) was validated again for all channels on this MC+LE-CR, and was found to provide a good description for invariant masses above 0.4~\TeV, giving $\chi^2/\text{ndf}\simeq 1.37, 0.97, 1.70, 1.13$ for the $\mjjl$, $\mjjll$, $\mjbl$ and $\mbbl$ distributions respectively, where ndf is the number of degrees of freedom. The fit results of these four MC+LE-CR distributions are shown in
Figure~\ref{fig:mjjl:signal:mc_with_fit}. The distributions of the fit residuals are also consistent with normal distributions with means of zero.
The fit parameters are strongly constrained by the low-mass region, below 1~\TeV, which has the most events.
Functions with the same form as Eq.~(\ref{eq:function}) but with fewer than five parameters fail to describe $\mjjl$ adequately in the LE-CR and MC+LE-CR regions.
 
An alternative five-parameter function for the description of the background was investigated. This alternative function, proposed in Ref.~\cite{Aad:2020kep}, differs from Eq.~(\ref{eq:function}) by replacing $p_5\ln^{2}x$ with $p_5/\sqrt{x}$. Similarly to the nominal function, applying the alternative function to the LE-CR and MC+LE-CR regions leads to residuals distributed according to a normal distribution with a mean consistent with zero. Differences between fits of the $\mjjl$ and $\mjjll$ distributions with the nominal and alternative functions were statistically insignificant. As the two functions were statistically equivalent when applied to the MC+LE-CR, Eq.~(\ref{eq:function}) was chosen as the nominal function in order to remain consistent with Ref.~\cite{Aad:2020kep}.

To investigate potential biases in the description of the MC+LE-CR by
Eq.~(\ref{eq:function}), `signal-injection' and `spurious-signal' tests are performed.
For the signal-injection test, signal events modelled according to Gaussian
distributions are added to the expected background distribution to assess
whether or not the correct numbers of events can be extracted using
signal-plus-background fits with unconstrained fit parameters,
assuming the known Gaussian signal shape.
The signal-injection tests are performed for the $\mjjl$, $\mjjll$, $\mjbl$ and $\mbbl$ distributions, assuming injected signals with various widths from 3\% to 15\% of the mean mass. The signal-injection efficiency correction factor is defined as the ratio of the extracted events above the $5p$ fit to the number of injected events. These factors are applied to the limits for each mass spectra to be presented. The correction factors are small and their values range from 0.96 to 1.0.
 
For the spurious-signal test, signal-plus-background fits are run on the background-only
spectra of the control region for different signal masses, and the extracted signal yield
is taken as an estimate of a potentially false signal.
The extracted signal yield in the spurious-signal tests is less than 50\% of the statistical
uncertainty in each mass bin.
This event rate for the signal region is considered as a source of
systematic uncertainty in the estimation of cross-section limits (discussed in Section~\ref{sec:sys}).
 
Based on the studies of the control regions, the background-only hypothesis for the signal region is constructed using Eq.~(\ref{eq:function}), over the mass range starting from 0.4~\TeV.
 
To determine whether the data deviate significantly from the
background-only hypothesis, the BumpHunter (BH) algorithm \cite{bumphunter} is used.
This test calculates the significance of any excess found in  mass
intervals at all possible locations in the binned invariant mass.
The width of the search window varies from a minimum of two
bins to half of the full width of the invariant mass distribution.
For each of the invariant mass intervals, a local $p$-value is calculated from a
hypothesis-test statistic. The look-elsewhere effect~\cite{Gross:2010qma} is taken into account by combining the separate
hypothesis tests to form a new hypothesis test, and calculating the minimum $p$-value amongst all tests.
A global $p$-value is then calculated and transformed into a significance by assuming that bin-by-bin
fluctuations of the data follow a Poisson distribution.
Pseudo-experiments are then used to determine the most significant local excess and, finally, a global significance is calculated.
Signal-injection tests show a lower signal-extraction efficiency for wide signals than for  narrow signals.

If no statistically significant deviations are observed according to the BH test, upper limits are set at the 95\% confidence level (CL) on the production cross-section times branching ratio using the Bayesian method  \cite{Caldwell:2008fw}
for generic resonances, as well as for the benchmark BSM models discussed in Section~\ref{sec:mc}.
Systematic uncertainties are included as nuisance parameters, assuming Gaussian priors.
A constant prior is used for the signal normalisation.
The $0.95$ quantile in the signal contribution is calculated;
this is taken as the upper limit on the number of possible signal events in the data, for each
resonance mass and width hypothesis. This value, divided by the integrated luminosity,
provides a measure of the upper limit on the cross-section times acceptance times efficiency times
branching ratio for a resonance with that mass and width.
This method is also used to determine the expected limits, along with the corresponding one-standard-deviation ($1\sigma$) and two-standard-deviation  ($2\sigma$) uncertainty bands, using 100 pseudo-experiments.
 
For the BSM models, the limits calculated using the reconstructed events are corrected for the acceptance and efficiency as a function of the mass.
The acceptance is constrained by the $\pt$ and $\eta$ requirements placed on the leptons and jets, and by the minimum dijet invariant mass used in this analysis.
The acceptance is typically 50\% for the lowest mass point, and increases to 60\%--90\% for the highest mass,
depending on the model.
The efficiency correction accounts for various instrumental effects, such as the trigger,
lepton identification, and lepton reconstruction inefficiencies.
A typical efficiency, averaged over the two lepton flavours, is 65\%--75\%,
depending on the particle mass and the type of BSM model.
 
\section{Systematic uncertainties}
\label{sec:sys}
 
A number of experimental systematic uncertainties are considered for the limit calculations.
They are divided into signal-shape-changing uncertainties that affect the limit calculations, uncertainties in the background shape, and
uncertainties in efficiencies (and acceptances) that
affect the overall normalisation of
the final limits for the BSM benchmark models. The first two categories of uncertainties are common to the limit calculations for
Gaussian signals and  BSM models.
Systematic uncertainties affecting the shapes
of invariant mass distributions  are estimated from the BSM MC event samples. They are included as nuisance parameters in the calculation of the Bayesian limits. The uncertainty of 1.7\% in the integrated luminosity is used for all limits.
 
The effects of jet energy scale (JES) and jet energy resolution (JER)
uncertainties~\cite{ATLAS:2020cli} are estimated using signal model MC events.
The combined effect of all the systematic uncertainties is
to raise the cross-section upper limits relative to the limits without uncertainties.  This
effect  has a typical size comparable to the width of the $1\sigma$ statistical uncertainty band around the expected limits. Uncertainties related to $b$-jet tagging are included in the
$\mjbl$ and $\mbbl$ invariant masses.
The JES uncertainty accounts for more than 50\% of the total systematic uncertainty, while the JER uncertainty makes the second-largest contribution. The effect of lepton energy scale and resolution uncertainties is found to be negligible. The effects of lepton trigger, identification and reconstruction uncertainties, contributing to differences between data and MC events, are taken into account by assigning a combined  uncertainty of 1\%.
 
For the generic Gaussian signals, JES and JER systematic uncertainties from the $W'\to W Z' \to \ell \nu \> q\bar{q}$ MC
events are used, parameterised as a function of mass to generate
uncertainties for mass values that were not used in generating the $W'\to W Z'$ samples.
As a cross-check, other models predicting different signal widths are also used, but no significant
differences are observed.
 
The uncertainties arising from imperfect knowledge of the background shapes are
investigated by using the alternative background fit and the spurious-signal test (discussed previously) using the nominal $5p$ fit to the data.
The uncertainty due to the choice of background function is estimated by performing the
background fit with the alternative fit function. The difference between the results of this fit and the nominal one, averaged across a set of pseudo-data, is considered as a systematic uncertainty.  Both uncertainties are found to have a negligible effect on the limits.
 
Uncertainties in the efficiency and acceptance correction factors related to
JES, JER, $b$-jet and lepton identification (and reconstruction) are found to be smaller than $1\%$.
An additional uncertainty in the limits is associated with the PDF choice for the acceptance calculation \cite{Aaboud:2017yvp}. This is accounted for by assigning a 1\% PDF uncertainty to the calculated limits.


\section{Results}
\label{sec:result}
 
\begin{figure}
\begin{center}
\includegraphics[width=0.7\textwidth]{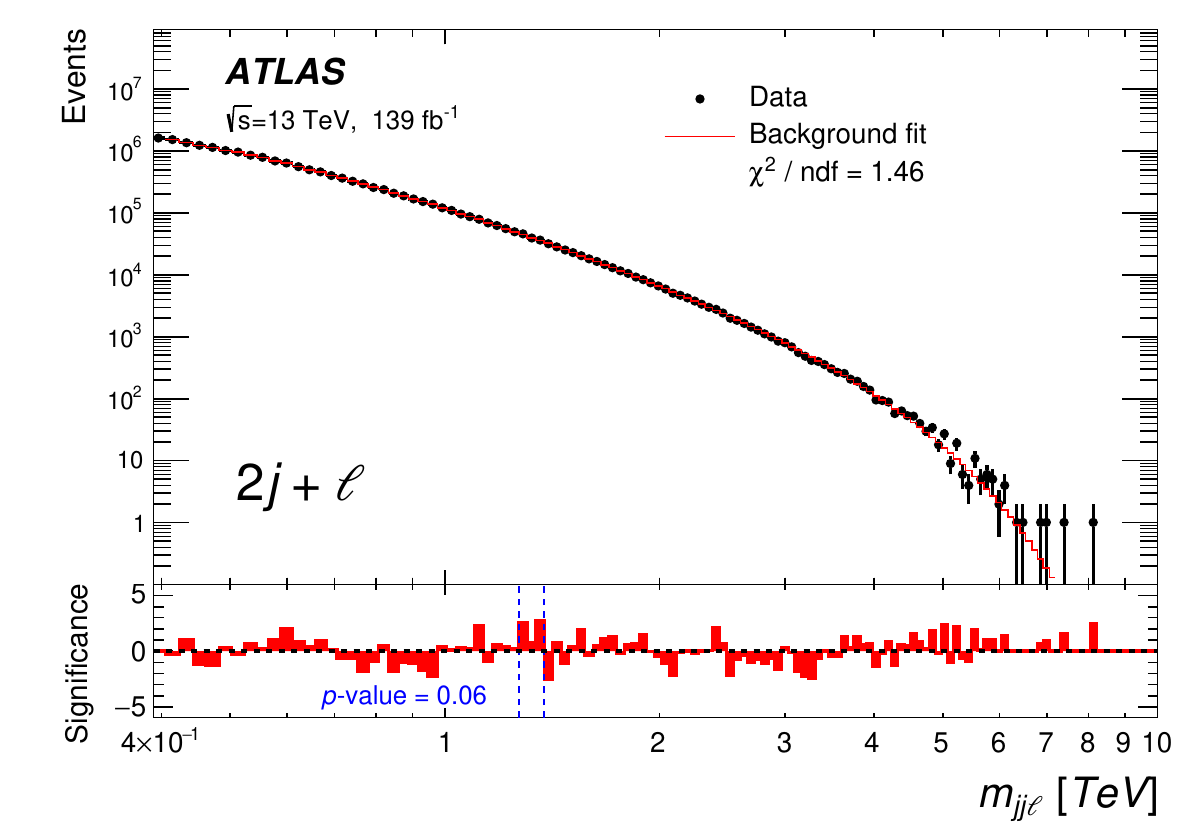}
\end{center}
\caption{Invariant $2j+\ell$ mass distribution for $\mjjl>0.4$~\TeV\ together with the fit with the $5p$ background function.
The lower panel shows the bin-by-bin significances of deviations from the background hypothesis. The largest deviation reported by the BumpHunter algorithm is shown by the vertical dashed lines. The global $p$-value of this deviation in the range 1.27--1.38~\TeV\ is indicated.}
\label{fig:pub_mjjl}
\end{figure}

\begin{figure}
\begin{center}
\includegraphics[width=0.7\textwidth]{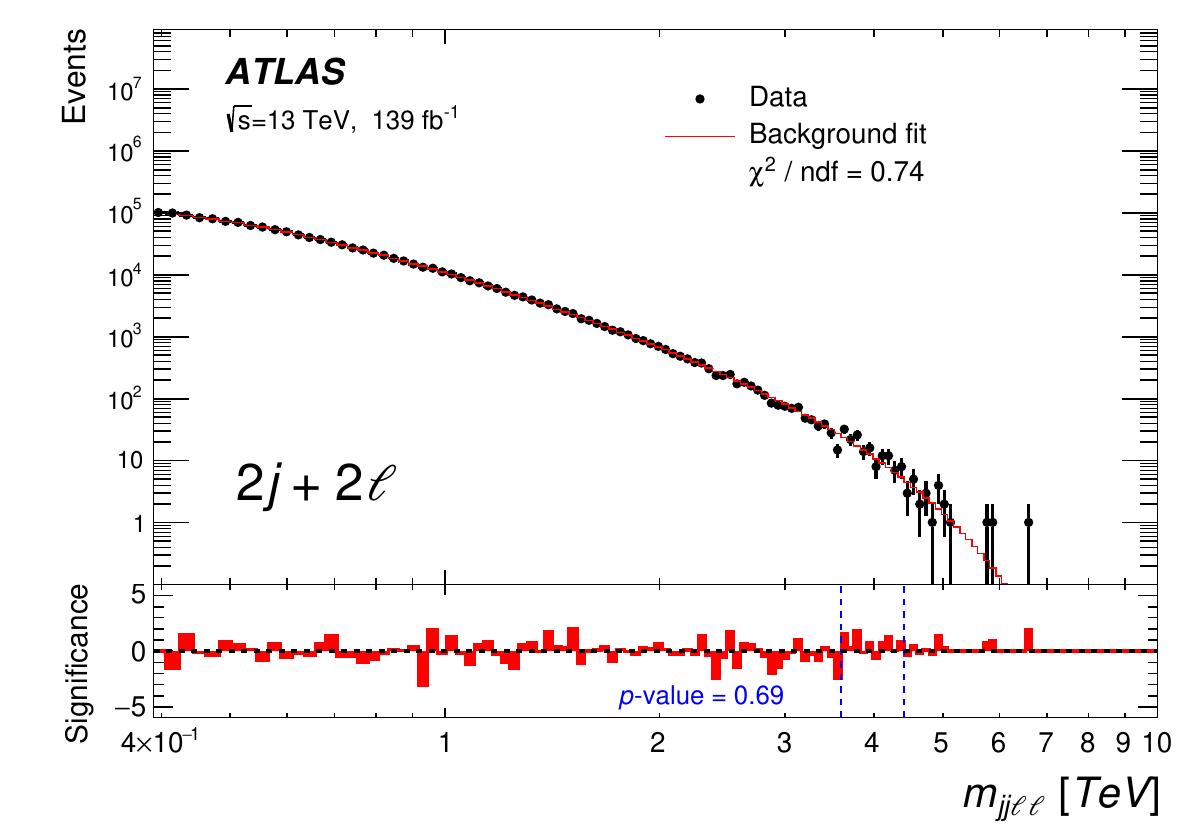}
\end{center}
\caption{Invariant $2j+ 2\ell$ mass distribution for $\mjjll>0.4$~\TeV\ together with the fit with the $5p$ background function.
The lower panel shows the bin-by-bin significances of deviations from the background hypothesis. The largest deviation reported by the BumpHunter algorithm is shown by the vertical dashed lines. The global $p$-value of this deviation in the range 3.59--4.41~\TeV\ is indicated.}
\label{fig:pub_mjjll}
\end{figure}

\begin{figure}
\begin{center}
 
\subfloat[$\mjbl$ invariant mass]{
\includegraphics[width=0.7\textwidth]{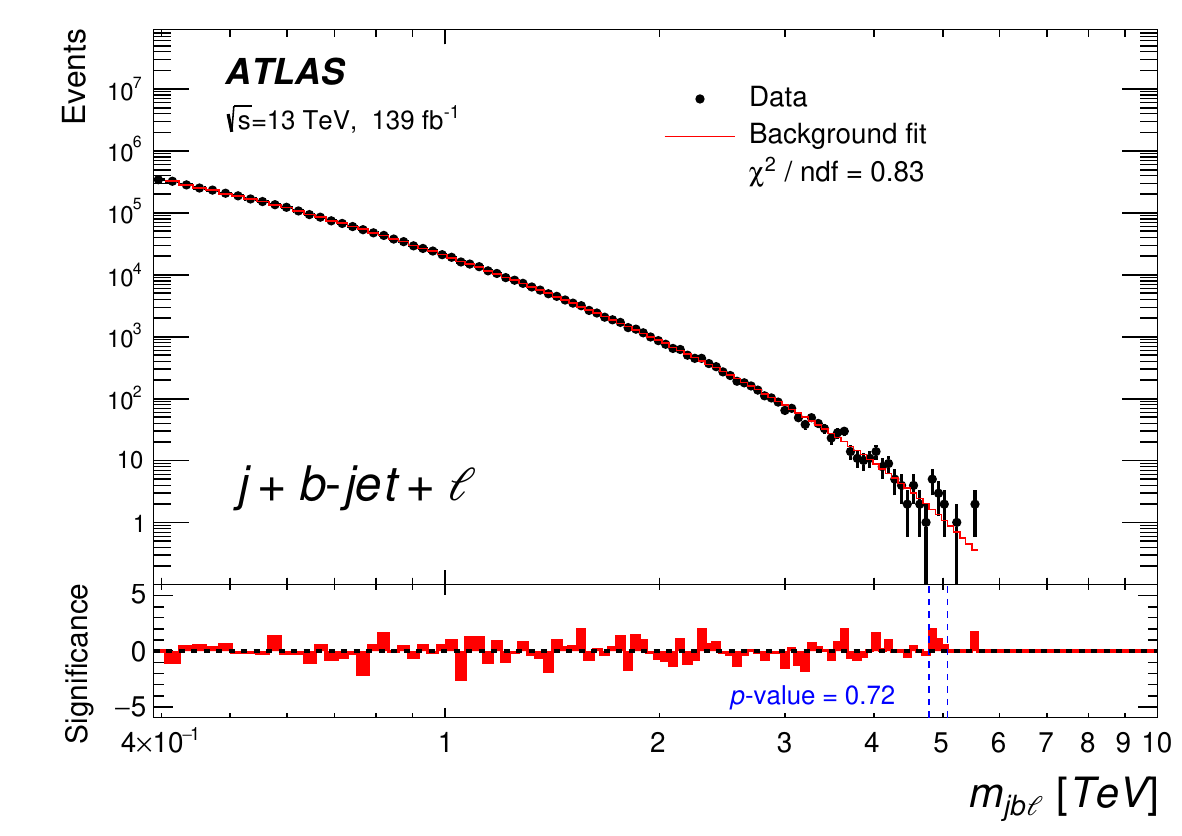}\hfill
}
 
\subfloat[$\mbbl$ invariant mass]{
\includegraphics[width=0.7\textwidth]{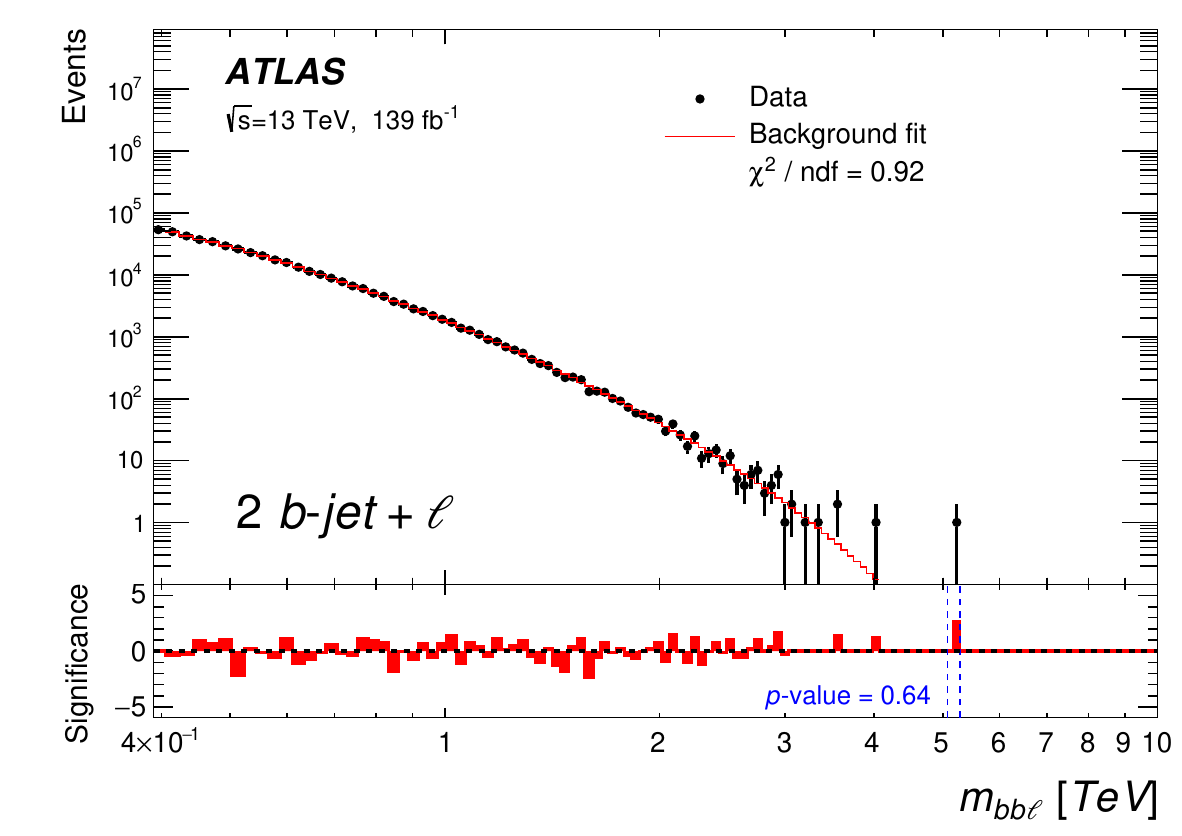}
}

\end{center}
\caption{Invariant mass distributions (a) $\mjbl$ and (b) $\mbbl$  with the $5p$ background fits. The lower panels show the bin-by-bin significances of deviations from the background hypothesis. The largest deviations reported by the BumpHunter algorithm are indicated by the vertical dashed lines. The global $p$-values of these deviations in the range 4.78--5.07~\TeV\ (for $\mjbl$) and 5.07--5.27~\TeV\ (for $\mbbl$) are indicated.}
\label{fig:pub_mjbl}
\end{figure}

Figures~\ref{fig:pub_mjjl}, \ref{fig:pub_mjjll} and \ref{fig:pub_mjbl}  show the
invariant mass distributions discussed above, together with the
results of the likelihood fits using the $5p$ function from Eq.~(\ref{eq:function}).
The quality of the fit in terms of the $\chi^2/$ndf is indicated. In order to establish the presence or absence of a signal in the $\mjjl$ distribution, the BH algorithm is used.  It is also used to calculate
local and global significances of the largest deviation from the background hypothesis.

Figure~\ref{fig:pub_mjjl} shows the invariant mass distribution $\mjjl$ obtained from the selected $2j+\ell$ events.
The data are described by the $5p$ fit function, with $\chi^2=149$ and $\text{ndf}=102$. The lower panel shows the significances \cite{Choudalakis_2012} of deviations from the background hypothesis, which can be approximated by $(d_i - f_i)/\delta_i$, where $d_i$ is
the number of events in the bin $i$, $f_i$ is the fit value, and $\delta_i$ is an uncertainty.
This uncertainty includes the statistical and systematic uncertainties of the fit result $f_i$.
 
The largest deviation of the $\mjjl$
data from the background-only hypothesis reported by the BH is near 1.3~\TeV, with a local $p$-value of $2\times 10^{-5}$,
corresponding to a significance of $3.5$ standard deviations.
Accounting for the look-elsewhere effect for the $\mjjl$ distribution, the global $p$-value for the largest deviation using the nominal fit is $0.06$, leading to a global significance of $1.5\sigma$.
When using the alternative background fit, this deviation  has a lower local significance of $2.8\sigma$.
This deviation from the background hypothesis is thus consistent with a statistical fluctuation.
 
The deviation near 1.3~\TeV\ is equally present in the electron and muon channels, but its statistical
significance is smaller than for the combined channel.  For the separate channels, the BH reports that
the largest deviations are in the region near 5~\TeV.
The global $p$-values for the  deviations near 5~\TeV\ are 0.13 (for muons) and 0.38 (for electrons).
Thus, such deviations from the $5p$ background are also consistent with  statistical fluctuations.
 
Figure~\ref{fig:pub_mjjll} shows the $\mjjll$ mass distribution for the combined muon and electron channels.
The global $p$-value for the largest deviation, near 4~\TeV, is $0.69$, indicating
no significant deviations from the $5p$ background. The separate muon and electron channels do not show
significant deviations.
 
Figure~\ref{fig:pub_mjbl} shows the distributions for three-body masses with one or two $b$-jets.
The global $p$-values for the largest deviations are larger than 0.6.
Thus the distributions do not show significant deviations from the background.

\subsection{Model-independent limits}
 
In the absence of any  statistically significant deviations from the $5p$ background function, the Bayesian method is used to set 95\% CL upper limits on the cross-section for new processes that have a signature of a new particle with decays producing two jets and leptons.
The limits are  obtained for hypothetical signals with a Gaussian shape.\footnote{Using the Gaussian limits requires some care in treating the realistic shapes from specific models. For example, the appendix of  Ref.~\cite{ATLAS:2014ktg} gives a recipe for how the Gaussian limits should be used for dijet studies.}
 
\begin{figure}[ht]
\begin{center}
\includegraphics[width=0.6\textwidth]{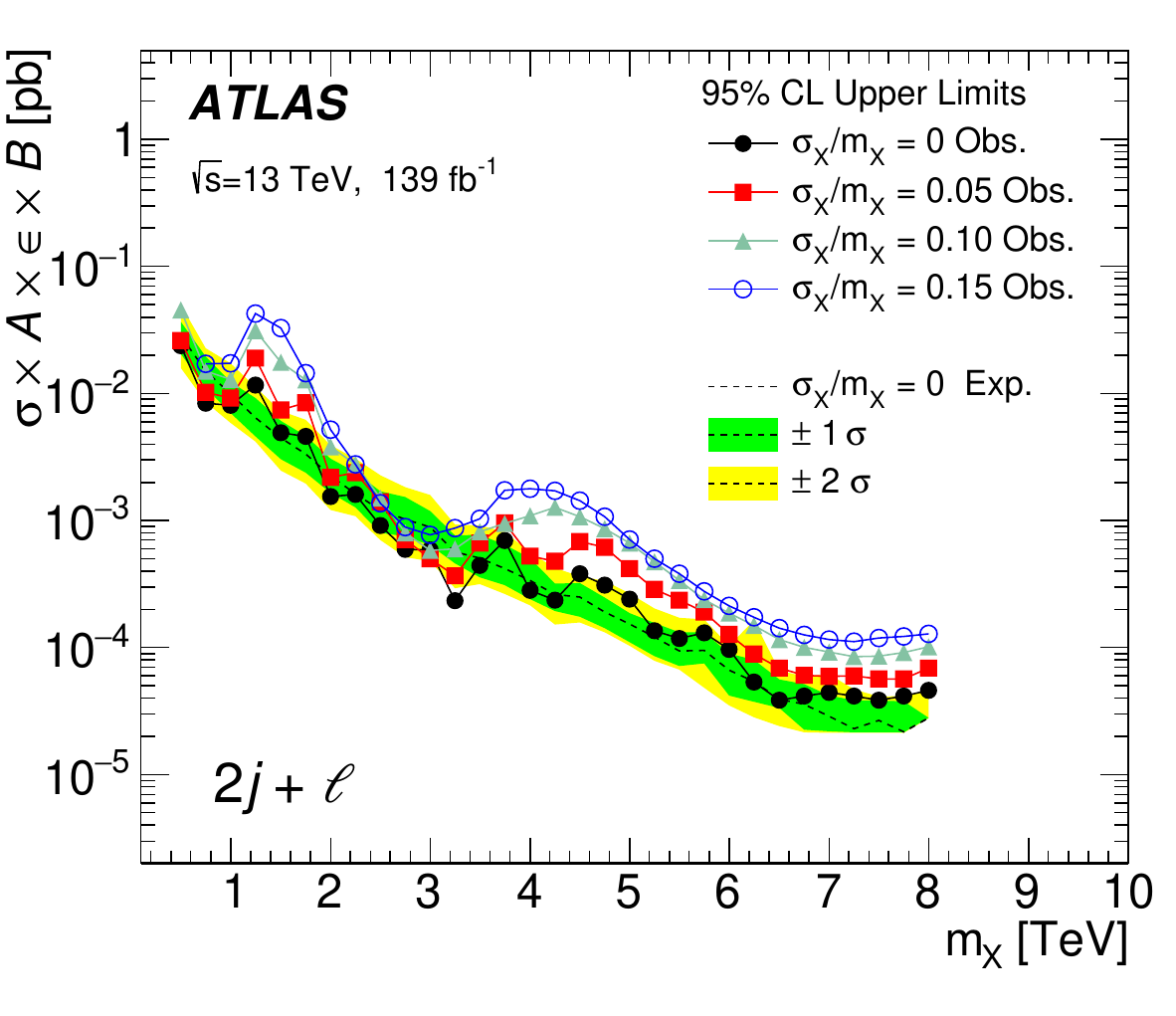}
\end{center}
\caption{
The 95\% CL observed upper limits, in the $\mjjl$ channel, on the cross-section times acceptance ($A$), efficiency
($\epsilon$) and branching ratio ($B$) for a  Gaussian signal with various signal widths.
The expected limit and the corresponding
$\pm 1\sigma$ and $\pm 2\sigma$ bands are shown for $\sigmaX/\mX=0$ signals.
The limit for a width of 15\% is shown starting from 0.75~\TeV\ in order to avoid the region close to the minimum value of the $\mjjl$ distribution.
}
\label{fig:gauss_all:mjjl}
\end{figure}
 
\begin{figure}[ht]
\begin{center}
\includegraphics[width=0.6\textwidth]{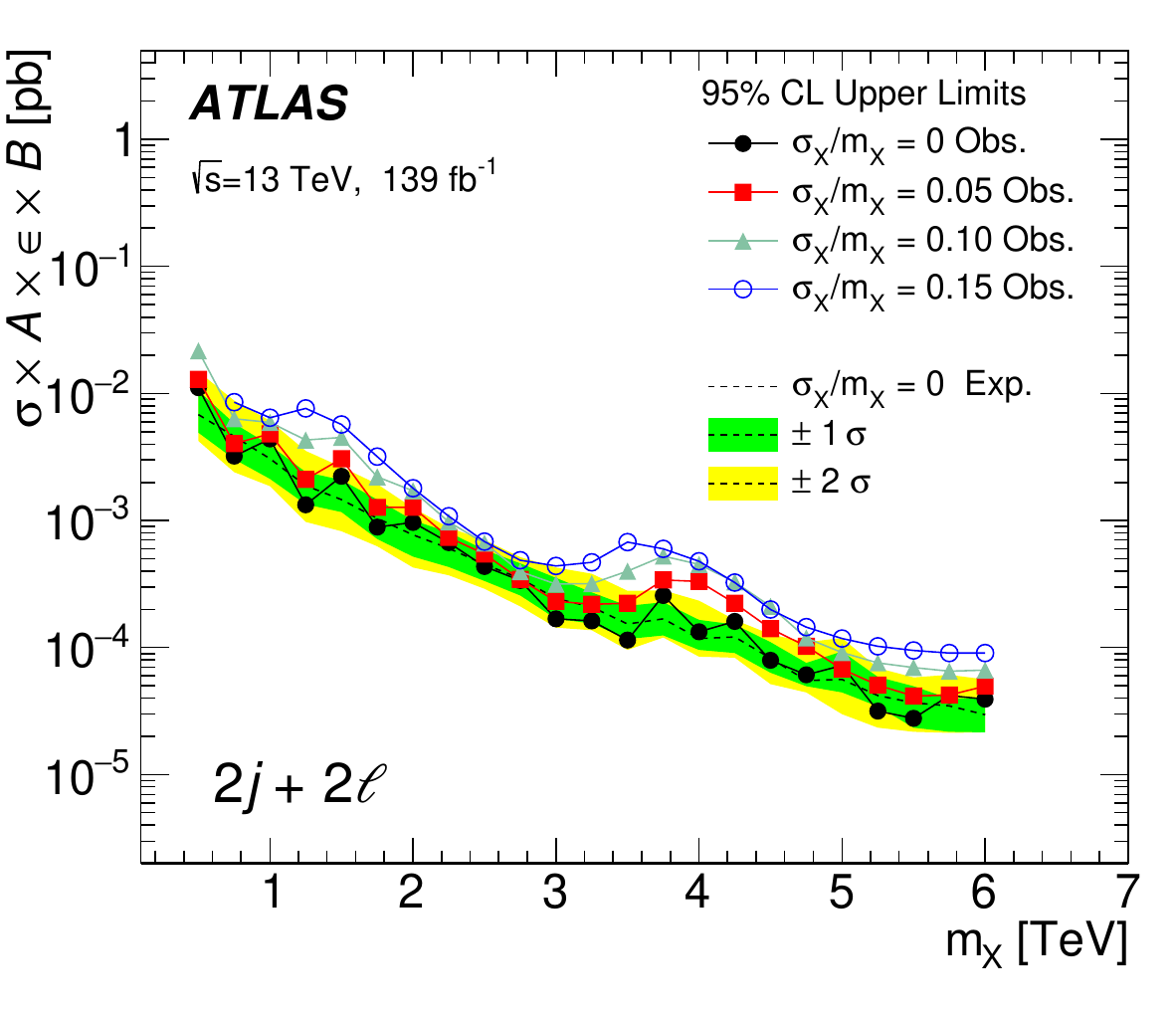}
\end{center}
\caption{
The 95\% CL observed upper limits, in the $\mjjll$ channel, on the cross-section times acceptance ($A$), efficiency
($\epsilon$) and branching ratio ($B$) for a  Gaussian signal with various signal widths.
The expected limit and the corresponding
$\pm 1\sigma$ and $\pm 2\sigma$ bands are shown for $\sigmaX/\mX=0$ signals.
The limit for a width of 15\% is shown starting from 0.75~\TeV\ in order to avoid the region close to the minimum value of the $\mjjll$ distribution.
}
\label{fig:gauss_all:mjjll}
\end{figure}
 
Figure~\ref{fig:gauss_all:mjjl} shows the $95\%$ CL expected and observed limits on the cross-section
times acceptance ($A$), signal efficiency ($\epsilon$) and branching ratio ($B$) of a BSM particle
decaying into two jets and a lepton.
The observed limits are shown for different widths, ranging from a narrow width
-- comparable to the experimental resolution of about 3\% -- up to 15\% of the mean mass. The expected limit and
the corresponding $\pm 1\sigma$ and $\pm 2\sigma$ bands are shown for narrow width ($\sigmaX/\mX=0$) resonances. As
expected, the limits are weaker for broader resonances that span  more $\mjjl$ bins.
Signals making Gaussian contributions to the $\mjjl$ distribution are excluded by the obtained limits as a
function of the mean mass. The expected limit and
the corresponding $\pm 1\sigma$ and $\pm 2\sigma$ bands are shown for the narrowest Gaussian width. The observed exclusion limits range from $20$~fb to $0.1$~fb in the mass range 0.5--8~\TeV.

\begin{figure}[ht]
\begin{center}
 
\subfloat[Limits for $\mjbl$]{
\includegraphics[width=0.6\textwidth]{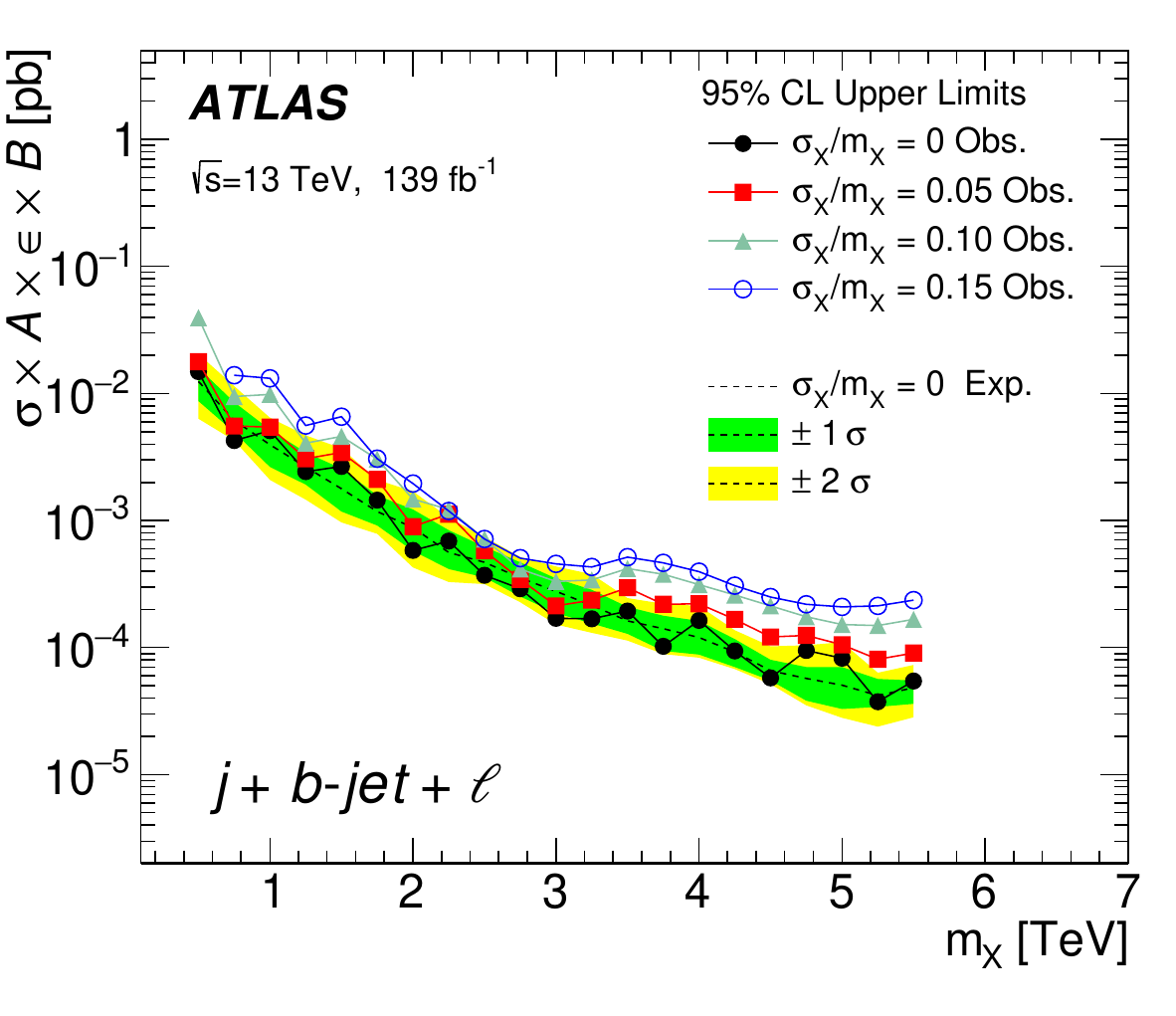}
}
 
\subfloat[Limits for $\mbbl$]{
\includegraphics[width=0.6\textwidth]{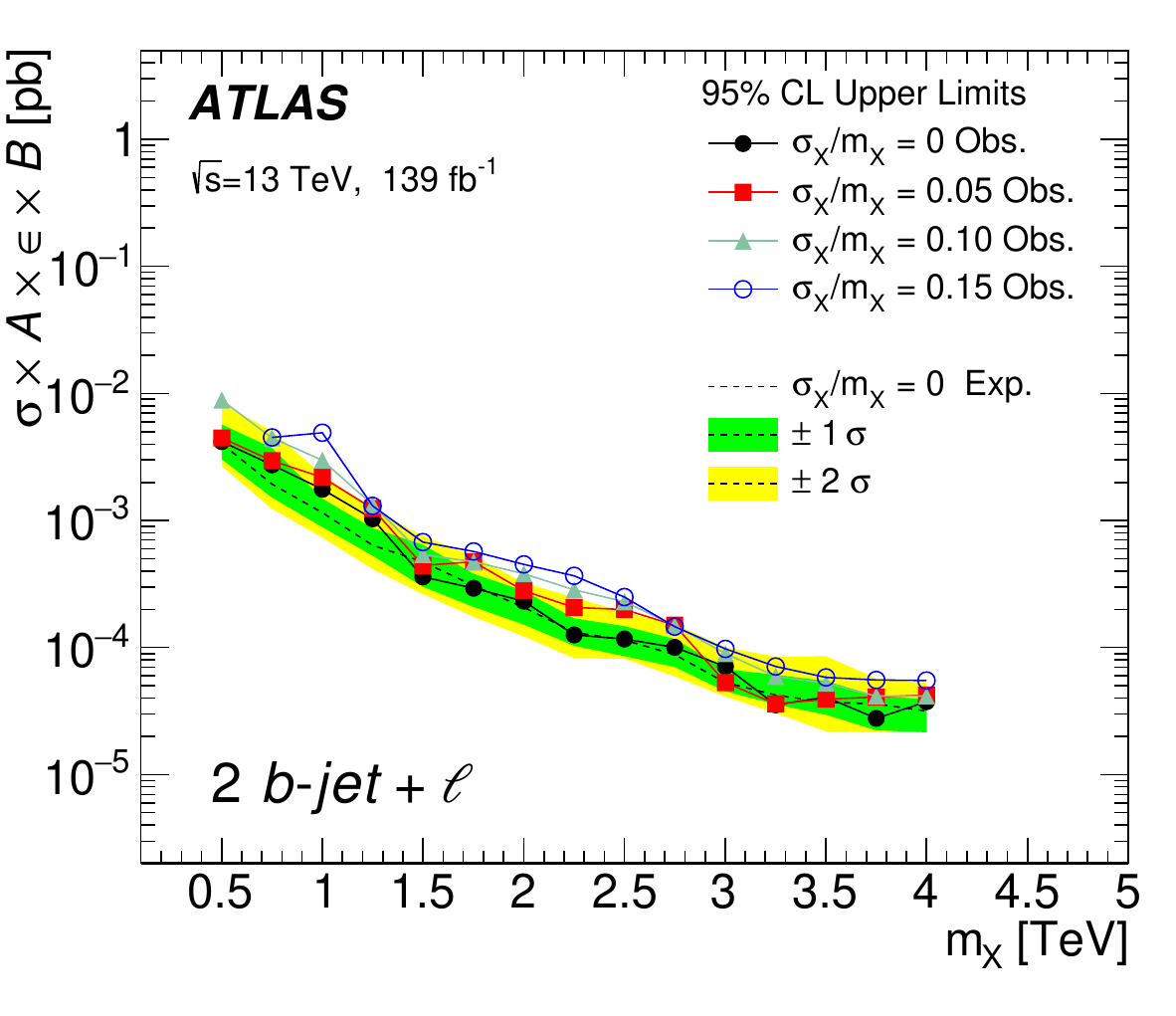}
}
 
\end{center}
\caption{
The 95\% CL observed upper limits, in the $\mjbl$ and $\mbbl$ channels, on the cross-section times acceptance ($A$), efficiency
($\epsilon$) and branching ratio ($B$) for a  Gaussian signal with various signal widths.
The expected limits and the corresponding
$\pm 1\sigma$ and $\pm 2\sigma$ bands are shown for $\sigmaX/\mX=0$ signals.
The limits for a width of 15\% are shown starting from 0.75~\TeV\ in order to avoid the region close to the minimum value of the invariant mass distributions.
}
\label{fig:gauss_all:mjbl}
\end{figure}

Figure~\ref{fig:gauss_all:mjjll}  shows the observed limits for four-body
invariant masses, $\mjjll$. The limits are obtained for
Gaussian signals with different widths, from a narrow width
-- compatible with the experimental resolution -- up to 15\% of the mean mass.
Figure~\ref{fig:gauss_all:mjbl}(a)  shows similar Gaussian
limits for the $\mjbl$ distribution.
Figure~\ref{fig:gauss_all:mjbl}(b)  shows the observed limits for the $\mbbl$ invariant mass.
The obtained limits range from about 10~fb to 0.1~fb for resonance masses between 0.5~\TeV\ and 4--5~\TeV.
 
\clearpage


\subsection{Limits on BSM models}

To calculate the exclusion limits for the benchmark models, the exact shapes of the $\mjjl$  and $\mjjll$ distributions of the signal MC events after simulation and
reconstruction are used. The Bayesian limits are calculated using the $\mjjl$ and $\mjjll$
distributions with the background description discussed earlier, and then the acceptance and efficiency corrections are applied. The limits on the composite-lepton model are calculated using $\mjjll$ in the muon channel for two masses, 250~\GeV\ and 500~\GeV, of a composite lepton $E$. The muon channel was chosen for this model as the $Z'$ preferentially couples to muons over electrons.
The calculated limits include the systematic uncertainties discussed in Section~\ref{sec:sys}.
 
\begin{figure}
\begin{center}
\subfloat[SSM model]{
\includegraphics[width=0.46\textwidth]{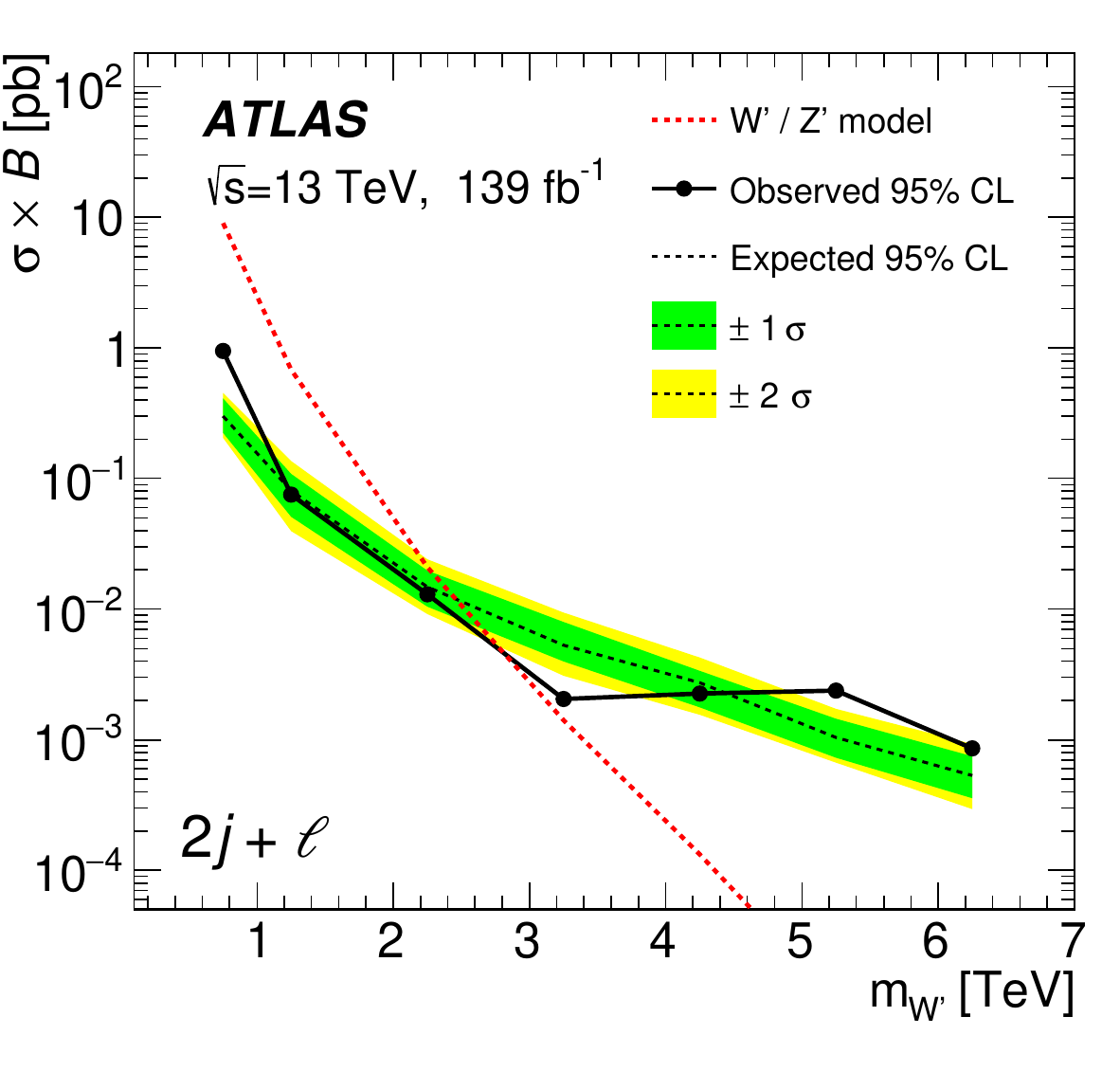}\hfill
}
\subfloat[$W_\text{KK}$ + $\phi$ model]{
\includegraphics[width=0.46\textwidth]{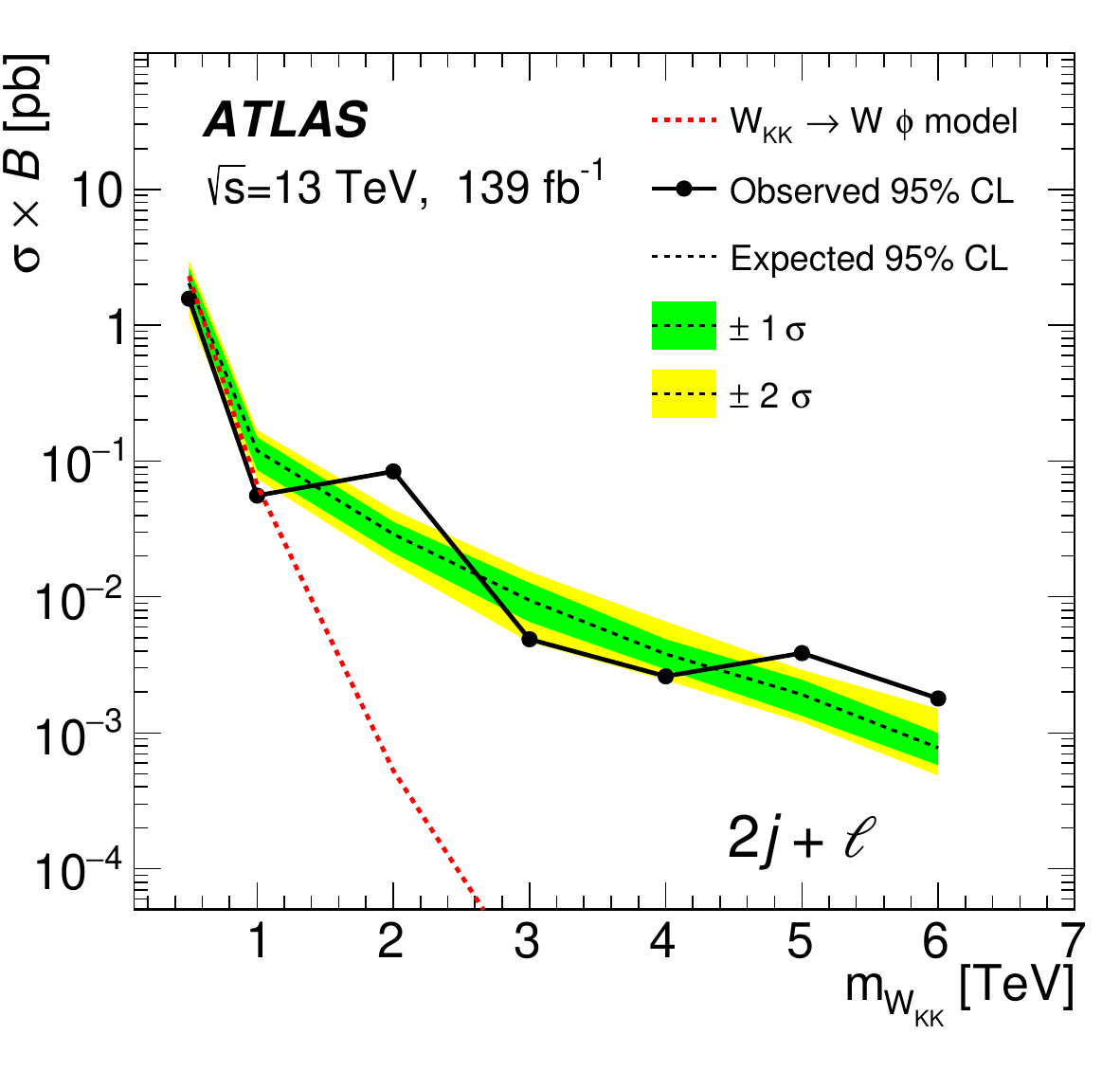}\hfill
}
 
\subfloat[$Z'W$ (DM) model]{
\includegraphics[width=0.46\textwidth]{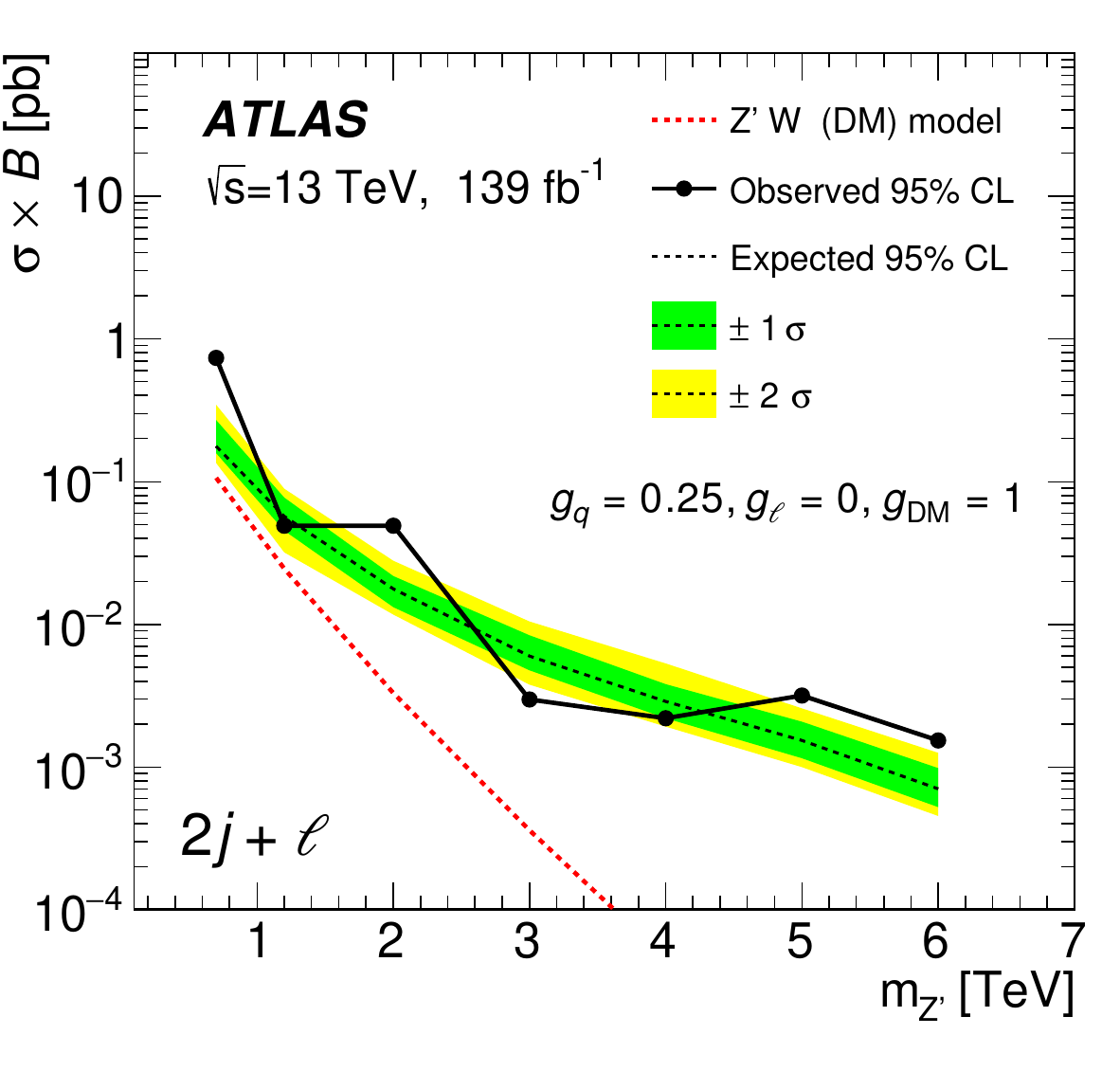}\hfill
}
\end{center}
\caption{
Observed (filled circles) and expected (dotted line with uncertainty bands)
95\% CL upper limits on the cross-section ($\sigma$)
times branching ratio ($B$) for
(a) the SSM $W'$ boson, (b) the $W_\text{KK}+\phi$ radion model with $m_{W_\text{KK}}-m_{\phi}=250$~\GeV, and (c) the $Z'W$ simplified DM model.
The presented limits were obtained using
the $\mjjl$ ($\ell=\mu, e$) distribution in events with at least one isolated lepton with $p_\text{T}^\ell>60$~\GeV.
Branching ratios of 100\% to the final states of interest were
assumed in the signal generation and computation of theory cross-sections. The black lines shown for the observed limits are provided to guide the eye.
}
\label{fig:limit_bsm_cor}
\end{figure}

\begin{figure}
\begin{center}
\subfloat[Composite-lepton model with $m_E=250$~\GeV]{
\includegraphics[width=0.46\textwidth]{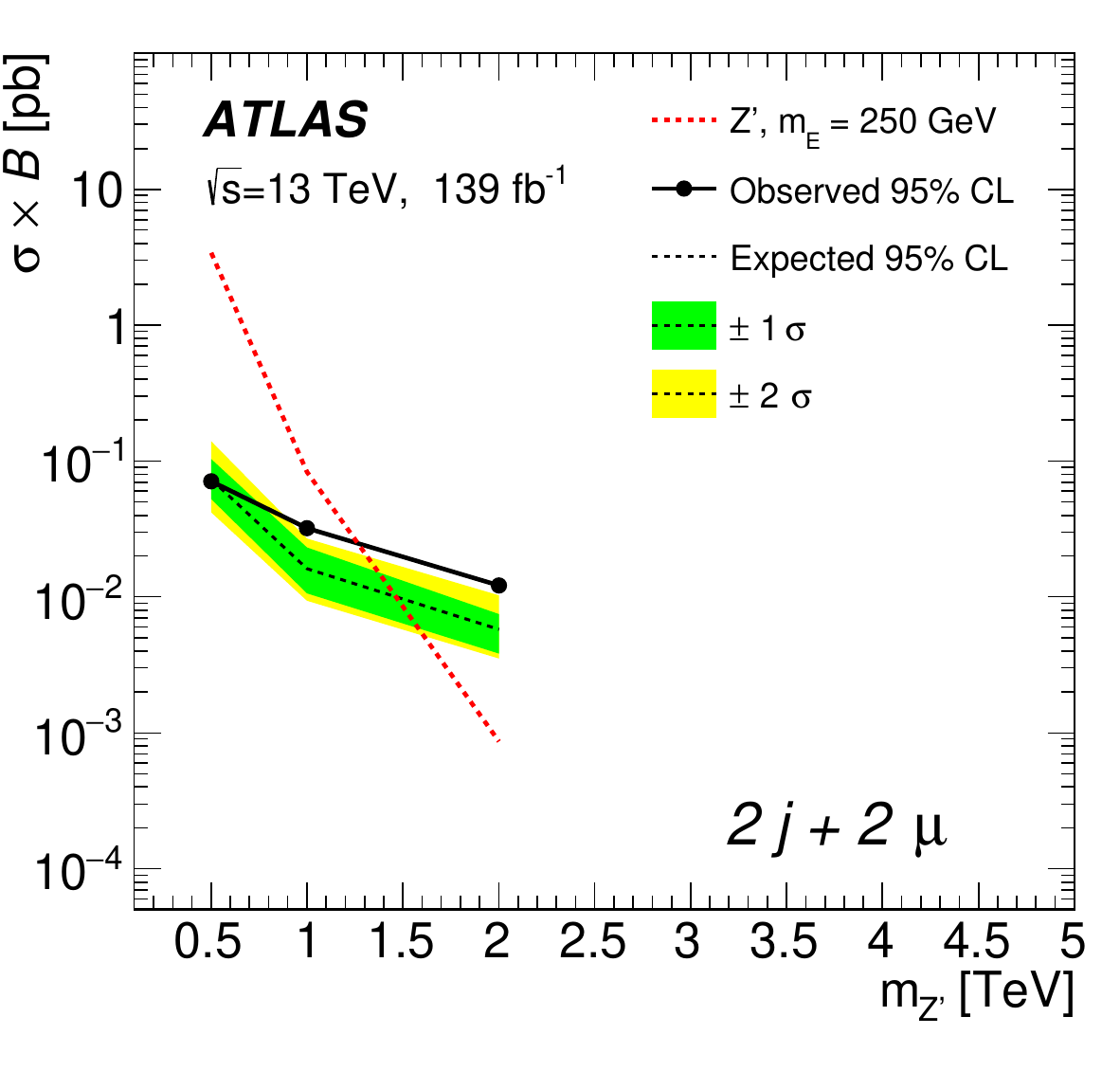}\hfill
}
\subfloat[Composite-lepton model with $m_E=500$~\GeV]{
\includegraphics[width=0.46\textwidth]{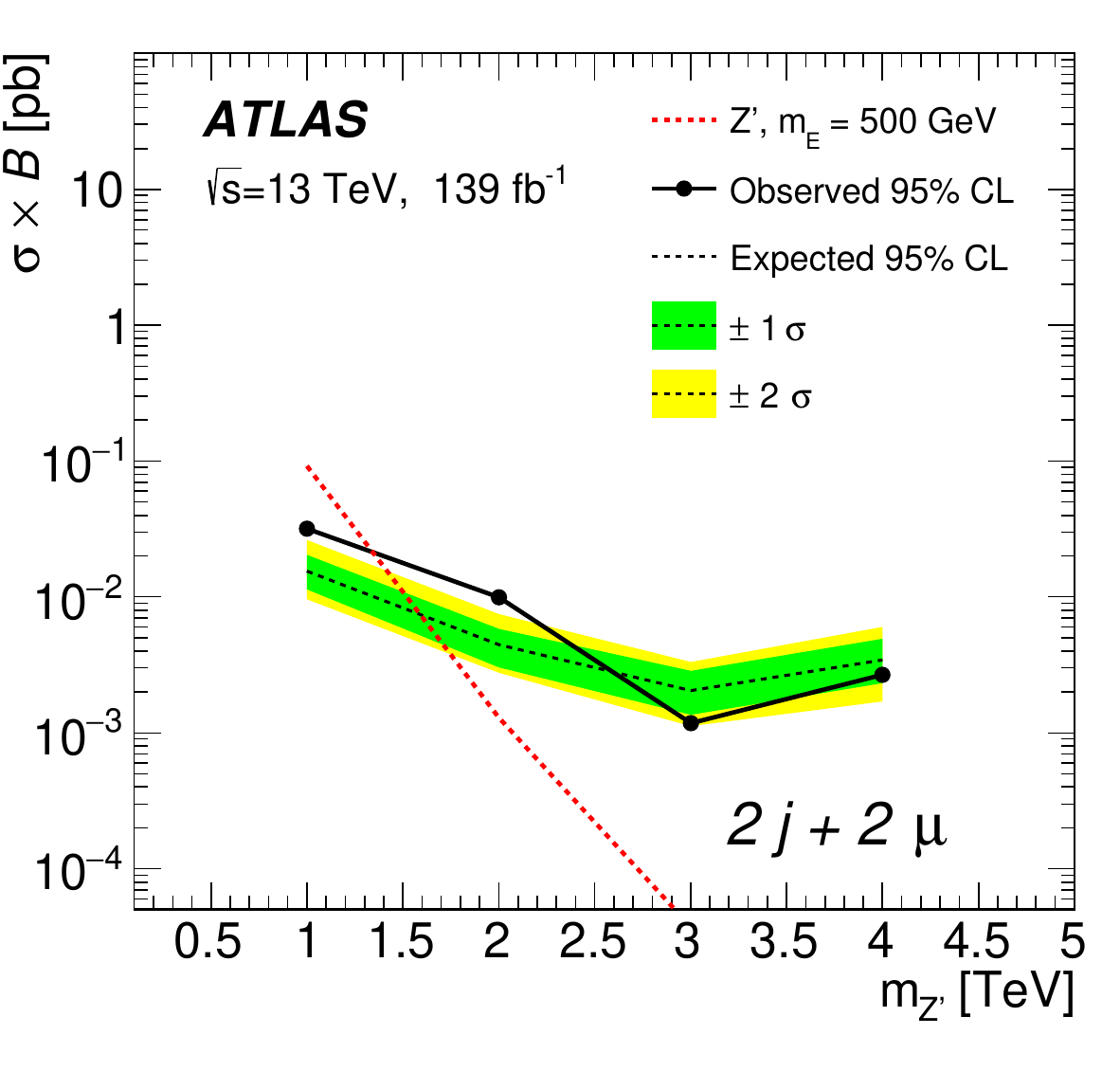}\hfill
}
 
\end{center}
\caption{
Observed (filled circles) and expected (dotted line with uncertainty bands)
95\% CL upper limits on the cross-section ($\sigma$)
times branching ratio ($B$) for $Z'\to E\mu$ where the composite lepton $E$ has a mass of (a) $250$~\GeV\ or (b) $500$~\GeV.
The presented limits were obtained using the $\mjjll$ ($\ell=\mu$) distribution in events with at least one isolated muon with $p_\text{T}^\ell>60$~\GeV.
Branching ratios of 100\% to the final states of interest were assumed in the signal generation and computation of theory cross-sections. The black lines shown for the observed limits are provided to guide the eye.
}
\label{fig:limit_bsm_com}
\end{figure}

Figures~\ref{fig:limit_bsm_cor} and \ref{fig:limit_bsm_com} show the observed (filled circles) and expected (dotted line with uncertainty bands) 95\%  CL upper limits on the cross-section times branching ratios.
These results exclude contributions from the $W'\to W Z'$ process in the
SSM for masses of the $W'$ boson (decaying into jet pairs) below
2.5~\TeV\ and a mass difference of 250~\GeV\ between the $W'$ and $Z'$ bosons, which
maximises the cross-section for this process. This is consistent with
the previous result for the same model using dijet events, where
a $Z'$ boson (decaying into jet pairs) with a mass below 2~\TeV\ was excluded \cite{Aad:2020kep}.
For the radion model, in which the mass-splitting between a radion $\phi$ and
$W_\text{KK}$ is small (250~\GeV), the data exclude  $\phi$  with a mass below 1~\TeV, as shown in Figure~\ref{fig:limit_bsm_cor}(b).
For the studied mass range, the current analysis cannot exclude the DM model with a $Z'$ axial-vector mediator, as seen in Figure~\ref{fig:limit_bsm_cor}(c).
Figure~\ref{fig:limit_bsm_com} shows the limits for the  composite-lepton model with a $Z'$ boson. The data exclude a $Z'$ boson with a mass below 1.3~\TeV\ for a composite lepton mass of 500~\GeV.

\FloatBarrier
\section{Conclusion}
A model-independent search for new resonances in invariant masses constructed from jets and leptons with $p_\text{T}^\ell>60$~\GeV\ using an integrated luminosity of \lumi~fb$^{-1}$ of proton--proton collisions with a
centre-of-mass energy of $\sqrt{s}=13$~\TeV\ recorded by the ATLAS detector at the LHC is presented.
Several three- and four-body invariant masses were studied:
$\mjjl$, $\mjjll$, $\mjbl$ and $\mbbl$ in the mass range from 0.4 to 8~TeV.
For the first time, searches in these invariant masses are performed with a data-driven technique, assuming a smoothly falling background described by a fit function, without relying on statistically limited MC predictions.
 
The multi-body invariant masses do not exhibit significant excesses
above a data-derived estimate of the smoothly falling background.
The most significant deviation from the data-derived estimate of the SM
background in the combined electron and muon channel is observed around
$\mjjl=1.3$~\TeV.  This excess
has a local statistical significance of 3.5 standard deviations.
After accounting for the look-elsewhere effect, this deviation corresponds
to a global statistical significance of $1.5\sigma$.
A check with an alternative description leads to a lower local significance of 2.8$\sigma$. Other multi-body invariant-mass combinations
show smaller local deviations.
Thus the data are consistent with the background-only hypothesis.
 
In the absence of signals indicating the presence of new phenomena, this analysis sets 95\% CL upper limits on the cross-section times acceptance, efficiency and branching fraction
for new processes that can
produce a Gaussian contribution to the invariant mass distributions of jets and leptons.
The limits are calculated for a Gaussian width ranging from that determined by the detector
resolution up to 15\% of the resonance mass.
These results exclude generic BSM models predicting resonances that decay to produce jets and leptons
with cross-sections larger than the limits reported in this paper. Under the assumption that a concrete BSM model can contribute to both the $\mjj$ and $\mjjl$ mass spectra, this study extends the search range by 2 TeV beyond that of previous model-independent searches \cite{Aad:2020kep} which did not include leptons in the invariant masses.
 
Model-dependent limits are also set on a variety of BSM models, without the use
of additional selection criteria tailored to the specific final states
investigated.
These results exclude contributions from the $W'\to W Z'$ process in the
SSM for masses of a $W'$ boson below 2.5~\TeV, assuming the mass difference
between the $W'$ and $Z'$ bosons is 250~\GeV\ to
maximise the cross-section for this process. The current analysis improves the SSM limits by 0.5 TeV compared to the dijet analysis \cite{Aad:2020kep}. For the radion model, in which the mass-splitting between a radion  $\phi$ and
$W_\text{KK}$ is small (250~\GeV), the data exclude  $\phi$  at masses below 1~\TeV.
For the composite-lepton model with a $Z'$ boson, the data exclude a $Z'$ boson with a mass below 1.3~\TeV\ for a composite lepton mass of 500~\GeV.


\section*{Acknowledgements}


We thank CERN for the very successful operation of the LHC, as well as the
support staff from our institutions without whom ATLAS could not be
operated efficiently.
 
We acknowledge the support of
ANPCyT, Argentina;
YerPhI, Armenia;
ARC, Australia;
BMWFW and FWF, Austria;
ANAS, Azerbaijan;
SSTC, Belarus;
CNPq and FAPESP, Brazil;
NSERC, NRC and CFI, Canada;
CERN;
ANID, Chile;
CAS, MOST and NSFC, China;
Minciencias, Colombia;
MEYS CR, Czech Republic;
DNRF and DNSRC, Denmark;
IN2P3-CNRS and CEA-DRF/IRFU, France;
SRNSFG, Georgia;
BMBF, HGF and MPG, Germany;
GSRI, Greece;
RGC and Hong Kong SAR, China;
ISF and Benoziyo Center, Israel;
INFN, Italy;
MEXT and JSPS, Japan;
CNRST, Morocco;
NWO, Netherlands;
RCN, Norway;
MEiN, Poland;
FCT, Portugal;
MNE/IFA, Romania;
JINR;
MES of Russia and NRC KI, Russian Federation;
MESTD, Serbia;
MSSR, Slovakia;
ARRS and MIZ\v{S}, Slovenia;
DSI/NRF, South Africa;
MICINN, Spain;
SRC and Wallenberg Foundation, Sweden;
SERI, SNSF and Cantons of Bern and Geneva, Switzerland;
MOST, Taiwan;
TENMAK, T\"urkiye;
STFC, United Kingdom;
DOE and NSF, United States of America.
In addition, individual groups and members have received support from
BCKDF, CANARIE, Compute Canada and CRC, Canada;
COST, ERC, ERDF, Horizon 2020 and Marie Sk{\l}odowska-Curie Actions, European Union;
Investissements d'Avenir Labex, Investissements d'Avenir Idex and ANR, France;
DFG and AvH Foundation, Germany;
Herakleitos, Thales and Aristeia programmes co-financed by EU-ESF and the Greek NSRF, Greece;
BSF-NSF and GIF, Israel;
Norwegian Financial Mechanism 2014-2021, Norway;
NCN and NAWA, Poland;
La Caixa Banking Foundation, CERCA Programme Generalitat de Catalunya and PROMETEO and GenT Programmes Generalitat Valenciana, Spain;
G\"{o}ran Gustafssons Stiftelse, Sweden;
The Royal Society and Leverhulme Trust, United Kingdom.
 
The crucial computing support from all WLCG partners is acknowledged gratefully, in particular from CERN, the ATLAS Tier-1 facilities at TRIUMF (Canada), NDGF (Denmark, Norway, Sweden), CC-IN2P3 (France), KIT/GridKA (Germany), INFN-CNAF (Italy), NL-T1 (Netherlands), PIC (Spain), ASGC (Taiwan), RAL (UK) and BNL (USA), the Tier-2 facilities worldwide and large non-WLCG resource providers. Major contributors of computing resources are listed in Ref.~\cite{ATL-SOFT-PUB-2021-003}.


\appendix

\clearpage
 
\printbibliography
 
\clearpage 
 
\begin{flushleft}
\hypersetup{urlcolor=black}
{\Large The ATLAS Collaboration}

\bigskip

\AtlasOrcid[0000-0002-6665-4934]{G.~Aad}$^\textrm{\scriptsize 101}$,
\AtlasOrcid[0000-0002-5888-2734]{B.~Abbott}$^\textrm{\scriptsize 119}$,
\AtlasOrcid[0000-0002-7248-3203]{D.C.~Abbott}$^\textrm{\scriptsize 102}$,
\AtlasOrcid[0000-0002-1002-1652]{K.~Abeling}$^\textrm{\scriptsize 55}$,
\AtlasOrcid[0000-0002-8496-9294]{S.H.~Abidi}$^\textrm{\scriptsize 29}$,
\AtlasOrcid[0000-0002-9987-2292]{A.~Aboulhorma}$^\textrm{\scriptsize 35e}$,
\AtlasOrcid[0000-0001-5329-6640]{H.~Abramowicz}$^\textrm{\scriptsize 150}$,
\AtlasOrcid[0000-0002-1599-2896]{H.~Abreu}$^\textrm{\scriptsize 149}$,
\AtlasOrcid[0000-0003-0403-3697]{Y.~Abulaiti}$^\textrm{\scriptsize 116}$,
\AtlasOrcid[0000-0003-0762-7204]{A.C.~Abusleme~Hoffman}$^\textrm{\scriptsize 136a}$
% ...
\bigskip
\\

% ...
$^{29}$Physics Department, Brookhaven National Laboratory, Upton NY; United States of America.\\
% ...
$^{35}$$^{(a)}$Facult\'e des Sciences Ain Chock, R\'eseau Universitaire de Physique des Hautes Energies - Universit\'e Hassan II, Casablanca;$^{(b)}$Facult\'{e} des Sciences, Universit\'{e} Ibn-Tofail, K\'{e}nitra;$^{(c)}$Facult\'e des Sciences Semlalia, Universit\'e Cadi Ayyad, LPHEA-Marrakech;$^{(d)}$LPMR, Facult\'e des Sciences, Universit\'e Mohamed Premier, Oujda;$^{(e)}$Facult\'e des sciences, Universit\'e Mohammed V, Rabat;$^{(f)}$Institute of Applied Physics, Mohammed VI Polytechnic University, Ben Guerir; Morocco.\\
% ...
$^{55}$II. Physikalisches Institut, Georg-August-Universit\"{a}t G\"ottingen, G\"ottingen; Germany.\\
% ...
$^{101}$CPPM, Aix-Marseille Universit\'e, CNRS/IN2P3, Marseille; France.\\
$^{102}$Department of Physics, University of Massachusetts, Amherst MA; United States of America.\\
% ...
$^{116}$Department of Physics, New York University, New York NY; United States of America.\\
% ...
$^{119}$Homer L. Dodge Department of Physics and Astronomy, University of Oklahoma, Norman OK; United States of America.\\
% ...
$^{136}$$^{(a)}$Departamento de F\'isica, Pontificia Universidad Cat\'olica de Chile, Santiago;$^{(b)}$Millennium Institute for Subatomic physics at high energy frontier (SAPHIR), Santiago;$^{(c)}$Instituto de Investigaci\'on Multidisciplinario en Ciencia y Tecnolog\'ia, y Departamento de F\'isica, Universidad de La Serena;$^{(d)}$Universidad Andres Bello, Department of Physics, Santiago;$^{(e)}$Instituto de Alta Investigaci\'on, Universidad de Tarapac\'a, Arica;$^{(f)}$Departamento de F\'isica, Universidad T\'ecnica Federico Santa Mar\'ia, Valpara\'iso; Chile.\\
% ...
$^{149}$Department of Physics, Technion, Israel Institute of Technology, Haifa; Israel.\\
$^{150}$Raymond and Beverly Sackler School of Physics and Astronomy, Tel Aviv University, Tel Aviv; Israel.\\
% ...
$^{a}$ Also Affiliated with an institute covered by a cooperation agreement with CERN.\\
$^{b}$ Also at An-Najah National University, Nablus; Palestine.\\
$^{c}$ Also at Borough of Manhattan Community College, City University of New York, New York NY; United States of America.\\
$^{d}$ Also at Bruno Kessler Foundation, Trento; Italy.\\
$^{e}$ Also at Center for High Energy Physics, Peking University; China.\\
$^{f}$ Also at Center for Interdisciplinary Research and Innovation (CIRI-AUTH), Thessaloniki; Greece.\\
% ...

\end{flushleft}


\end{document}